\documentclass[%
 reprint,
superscriptaddress,
 amsmath,
 amssymb,
 aps,
prb,
floatfix,
longbibliography
]{revtex4-1}

\usepackage{graphicx}
\usepackage{dcolumn}
\usepackage{bm}
\usepackage{color}
\usepackage{subfigure}

\begin{document}

\preprint{APS/123-QED}

\graphicspath{{mainfigures/}}
\title{Tunable band gap in twisted bilayer graphene}
\author{Xiu-Cai Jiang}
\email{These authors contribute equally to the paper.}
\affiliation{Shanghai Key Laboratory of Special Artificial Microstructure Materials and Technology, School of Physics Science and engineering, Tongji University, Shanghai 200092, P.R. China}
\author{Yi-Yuan Zhao}
\email{These authors contribute equally to the paper.}
\affiliation{Shanghai Key Laboratory of Special Artificial Microstructure Materials and Technology, School of Physics Science and engineering, Tongji University, Shanghai 200092, P.R. China}
\author{Yu-Zhong Zhang}
\email[Corresponding author.]{Email: yzzhang@tongji.edu.cn}
\affiliation{Shanghai Key Laboratory of Special Artificial Microstructure Materials and Technology, School of Physics Science and engineering, Tongji University, Shanghai 200092, P.R. China}

\date{\today}

\begin{abstract}
At large commensurate angles, twisted bilayer graphene which holds even parity under sublattice exchange exhibits a tiny gap. Here, we point out a way to tune this tiny gap into a large gap. We start from comprehensive understanding of the physical origin of gap opening by density functional theory calculations. We reveal that the effective inter-layer hopping, intra-layer CDW, or inter-layer charge imbalance favors a gap. Then, on the basis of tight-binding calculations, we suggest that a periodic transverse inhomogeneous pressure, which can tune inter-layer hoppings in specific regions of the moi$\rm\acute{r}$e supercell, may open a gap of over $100$~meV, which is further confirmed by first-principles calculations. Our results provide a theoretical guidance for experiments to open a large gap in twisted bilayer graphene.
\end{abstract}

\maketitle
\section{Introduction}
Electronic properties of twisted bilayer graphene (tBLG)~\cite{rozhkov2016electronic} are strongly affected by moi$\rm\acute{r}$e pattern and inter-layer coupling~\cite{PhysRevLett.109.186807}. As a result, many exotic phenomena such as Mott insulator~\cite{cao2018correlated,po2018origin}, superconductivity~\cite{cao2018unconventional,lu2019superconductors,balents2020superconductivity}, and higher-order topological insulator~\cite{park2019higher,kindermann2015topological} have been discovered in tBLG with different stacking configurations. Of particular interest, at large commensurate angles, tBLG, which holds even parity under sublattice exchange (SE), has not only graphene-like low-energy spectra~\cite{sprinkle2009first,shallcross2008quantum} but also an intrinsic gap at half filling~\cite{shallcross2008quantum,PhysRevB.81.165105}, making it a promising candidate for high-mobility field effect transistors. However, the gap is too small to switch off electrical conduction at room temperature, which impedes such applications~\cite{oostinga2008gate}.

Therefore, in order to overcome this problem, much effort has been made to investigate conditions for gap opening at half filling and properties of gapped states. It was proposed for the first time that the ground state of the tBLG with twisted angle $\theta=38.21^{\circ}$ is insulating with an intrinsic gap of less than 10 meV at half filling over a decade ago~\cite{shallcross2008quantum}. Using a long-wavelength theory, Mele pointed out that all SE-even structures are gapped, which are ascribed to pseudo spin-orbit coupling resulting from hoppings between layers~\cite{mele2010commensuration,mele2012interlayer}. From the tight-binding analyses, it was shown that gap only exists in SE-even tBLG with a twisted angle $\theta>\theta_{c}\approx1.89^{\circ}$~\cite{PhysRevB.92.075402}. Thus, the gap can be tuned by twisting different angles between two layers. Besides, in-plane strains can tune the gap as well, where a direct-indirect gap transition occurs under mixed in-plane strains~\cite{khatibi2019strain}. Moreover, recently, SE-even structures with large twisted angles were proposed to be higher-order topological insulators hosting corner states which emerge as long as the underlying symmetries are intact~\cite{park2019higher}. These findings further trigger the interest to study the gapped state in SE-even tBLG.

However, till now, despite extensive investigations, an effective way to open a large gap in SE-even tBLG has not yet been proposed. In fact, if only twisting is employed between layers, the gap is so small that one needs to control the twisted angle accurately in a clean sample to observe it at low temperature~\cite{rozhkov2017single}. Fortunately, the twisted angle can be controlled with a remarkable precision of $0.1^{\circ}$ in experiments~\cite{cao2016superlattice,kim2016van,kim2017tunable}. Besides, the techniques to compress the inter-layer distance down to $80\%$~\cite{zhao1989x}, to exert an in-plane tensile strain up to $6\%$ without inelastic relaxation~\cite{cao2020elastic}, and to apply a transverse electric field~\cite{liu2015observation} have been mastered. Applying these techniques to tBLG can lead to many novel phenomena, such as pressure dependence of flat bands~\cite{carr2018pressure,ge2021emerging,yndurain2019pressure}, magnetism induced by pressure at large twisted angle~\cite{yndurain2019pressure}, direct-indirect gap transition induced by mixed in-plane strains~\cite{khatibi2019strain}, and tunable gap induced by a transverse electric field~\cite{liu2015observation}. Furthermore, tBLG has intra-layer on-site potential differences~\cite{tepliakov2021crystal}, which are neglected in long-wavelength theory~\cite{mele2010commensuration} and tight-binding approximation~\cite{PhysRevB.92.075402}. On-site potential differences in the layer often induce an in-plane uneven charge distribution resulting in an intra-layer charge density wave (CDW), which may affect the gap. Hence, clarifying the effect of CDW on gap opening and finding a way to open a large gap in tBLG with above available techniques are topical subjects.

In this paper, we point out a way to tune the tiny gap of tBLG with twisted angle $\theta=38.21^{\circ}$ into a large gap. First, to show the tunable properties of gap, using first-principles calculations, we study the effects of external tuning parameters like a transverse homogeneous pressure, an in-plane biaxial tensile strain, or a transverse electric field on gap opening. We find that gap increases monotonously with increasing these tuning parameters. Second, to understand the reasons why gap increases with these parameters, we calculate effective inter-layer hopping and order parameters of intra- and inter-layer charge disproportionation. We reveal that enhancement of gap amplitude with applying pressures or strains is attributed to the increase of both intra-layer CDW and effective inter-layer hopping, where the latter always plays a dominant role on gap opening, while external electric field can induce an inter-layer charge imbalance, resulting in enlargement of the gap.  Third, using tight-binding model, we demonstrate how to open a larger gap by tuning inter-layer hoppings, and on-site potentials which can modify intra-layer CDW, in specific regions of the moi$\rm\acute{r}$e supercell, respectively. Finally, based on the guidance of tight-binding calculations, we suggest that a periodic transverse inhomogeneous pressure can open a gap of over $100$ meV, which is further confirmed by first-principle calculations.

Our paper is organized as follows. Sec.~\ref{TBLG:model_method} describe the details of the structure, the model, and the method we used.  Sec.~\ref{TBLG:results} present our main results, including gaps as functions of different external tuning parameters, density of state (DOS) and band structures, contour maps of difference of charge density, effective inter-layer hopping and order parameters of CDW states, analyses of tight-binding calculations and variations of different physical quantities as functions of effective periodic transverse inhomogeneous pressure.  Sec.~\ref{TBLG:DISCUSSION} includes a discussion of our results and Sec.~\ref{TBLG:conclusion} concludes with a summary.

\section{model and method}\label{TBLG:model_method}
The structure of tBLG with twisted angle $\theta=38.21^{\circ}$, which can be realized in experiment~\cite{koren2016coherent}, is shown in Fig.~\ref{fig:structure-and-gap}(a). According to the atomic environment, we identify three distinct regions in the moi$\rm\acute{r}$e supercell, namely, A, H, and M region as shown in Fig.~\ref{fig:gap-parameters}(a), corresponding to the region where atoms of two layers (green) sits on top of each other, the region where hexagon centers of two layers overlap (pink), and the region of the rests (blue), respectively.

To calculate the properties of tBLG under a transverse homogeneous pressure, an in-plane biaxial tensile strain, a transverse electric field, or a periodic transverse inhomogeneous pressure, density functional theory (DFT) calculations are employed, based on the projector augmented wave method~\cite{PhysRevB.50.17953} as implemented in the Vienna Ab initio Simulation Package (VASP)~\cite{PhysRevB.54.11169,kresse1996efficiency}. We choose the generalized gradient approximation (GGA) of Perdew-Burke-Ernzerhhor~\cite{PhysRevLett.77.3865} to the exchange-correlation potentials with van der Waals correction called vdW-DF2-B86r~\cite{PhysRevB.82.081101,PhysRevB.89.121103}, which is considered to be the functional with the best overall performance in multi-layer graphene~\cite{del2019layer}. A plane-wave energy cutoff of $450$ eV is employed. A $\Gamma$-centered K-point grid of 45$\times$45$\times1$ is used in the Brillouin-zone integral. The vacuum distance is set to be $20$ {\AA} to eliminate the coupling between periodic images of the layers in the direction perpendicular to the atomic planes.

To demonstrate how to open a large gap by tuning inter-layer hoppings and on-site potentials in distinct regions of the moi$\rm\acute{r}$e supercell, a simple model is introduced as
{\setlength\abovedisplayskip{0.1cm}
\setlength\belowdisplayskip{0.1cm}
\begin{eqnarray}\label{TB_Model}
H=H_{D}+\delta{H}_{t}+\delta{H}_{\Delta}.\label{eq:total hamiltonian}
\end{eqnarray}}
where
{\setlength\abovedisplayskip{0.1cm}
\setlength\belowdisplayskip{0.1cm}
\begin{equation}
\begin{aligned}
H_{D}&=
\sum\limits_{is\sigma}\Delta_{is}C_{is\sigma}^{\dag}C_{is\sigma}
-\sum\limits_{is\sigma}{\sum\limits_{js^{\prime}}}^{\prime}t_{isjs^{\prime}}C_{is\sigma}^{\dag}C_{js^{\prime}\sigma}\\
\end{aligned}
\end{equation}}
is the Hamiltonian derived from DFT calculations with tight-binding parameters obtained through transformation from Bloch space to maximally localized Wannier function basis by using WANNIER90 code~\cite{RevModPhys.84.1419,mostofi2008wannier90}. Totally, $28$ bands close to the Fermi energy are taken into account for each spin, which are mainly contributed from p$_z$ orbitals of $28$ carbon atoms in the moi$\rm\acute{r}$e supercell. $\delta{H}_{t}$ and $\delta{H}_{\Delta}$ are perturbed Hamiltonians that can be viewed as modelling of external perturbations like applying pressures, strains, and electric fields which result in inter-layer hoppings or on-site potentials deviating from their DFT values. Here, $\delta{H}_{t}$ contains five terms while $\delta{H}_{\Delta}$ includes three terms,
{\setlength\abovedisplayskip{0.1cm}
\setlength\belowdisplayskip{0.1cm}
\begin{equation}
\begin{aligned}
\delta{H}_{t}&=
\delta{H}_{t_A}+\delta{H}_{t_M^{1}}+\delta{H}_{t_M^{2}}+\delta{H}_{t_H^{1}}+\delta{H}_{t_H^{2}}\\
\delta{H}_{\Delta}&=
\delta{H}_{\Delta_A}+\delta{H}_{\Delta_M}+\delta{H}_{\Delta_H},
\end{aligned}
\end{equation}}
where
{\setlength\abovedisplayskip{0.1cm}
\setlength\belowdisplayskip{0.1cm}
\begin{equation}\label{TB_detail}
\begin{aligned}
\delta{H}_{t_A}&=
\delta{t}_{A}\sum\limits_{i\sigma}\sum\limits_{\langle{s,s^{\prime}}\rangle\in{A}}C_{is\sigma}^{\dag}C_{is^{\prime}\sigma}\\
\delta{H}_{t_M^{1}}&=
\delta{t}_M^{1}\sum\limits_{i\sigma}\sum\limits_{\langle{s,s^{\prime}}\rangle\in{M}}C_{is\sigma}^{\dag}C_{is^{\prime}\sigma}\\
\delta{H}_{t_M^{2}}&=
\delta{t}_M^{2}\sum\limits_{i\sigma}\sum\limits_{\langle{\langle{s,s^{\prime}}\rangle}\rangle\in{M}}C_{is\sigma}^{\dag}C_{is^{\prime}\sigma}\\
\delta{H}_{t_H^{1}}&=
\delta{t}_H^{1}\sum\limits_{i\sigma}\sum\limits_{\langle{s,s^{\prime}}\rangle\in{H}}C_{is\sigma}^{\dag}C_{is^{\prime}\sigma}\\
\delta{H}_{t_H^{2}}&=
\delta{t}_H^{2}\sum\limits_{i\sigma}\sum\limits_{\langle{\langle{s,s^{\prime}}\rangle}\rangle\in{H}}C_{is\sigma}^{\dag}C_{is^{\prime}\sigma}\\
\delta{H}_{\Delta_A}&=
\delta{\Delta}_A\sum\limits_{i\sigma}\sum\limits_{s\in{A}}C_{is\sigma}^{\dag}C_{is\sigma}\\
\delta{H}_{\Delta_M}&=
\delta{\Delta}_M\sum\limits_{i\sigma}\sum\limits_{s\in{M}}C_{is\sigma}^{\dag}C_{is\sigma}\\
\delta{H}_{\Delta_H}&=
\delta{\Delta}_H\sum\limits_{i\sigma}\sum\limits_{s\in{H}}C_{is\sigma}^{\dag}C_{is\sigma}.
\end{aligned}
\end{equation}}
Here, $i$ $(j)$,  $s$ $(s^{\prime})$, and $\sigma$ denote cell, sublattice, and spin index, respectively. $\langle{s,s^{\prime}}\rangle$ ($\langle{\langle{s,s^{\prime}}\rangle}\rangle$) means the summation over inter-layer nearest(next-nearest)-neighbor sites. $\delta{t}_{A}$, $\delta{t}_M^{1}$, $\delta{t}_M^{2}$, $\delta{t}_H^{1}$, and $\delta{t}_H^{2}$ are the deviations of inter-layer hopping integrals in corresponding regions (see Fig.~\ref{fig:gap-parameters}(a)) from their DFT values due to external perturbations, while $\delta{\Delta}_A$, $\delta{\Delta}_M$, and $\delta{\Delta}_H$ stand for the deviations of on-site potentials in corresponding regions from their DFT values. Thus, using this model, we can study the effect of external perturbations, which affect inter-layer hoppings or on-site potentials in distinct regions of the moi$\rm\acute{r}$e supercell, on gap opening.
\section{results}\label{TBLG:results}
\begin{figure}[htbp]
\includegraphics[width=0.47\textwidth]{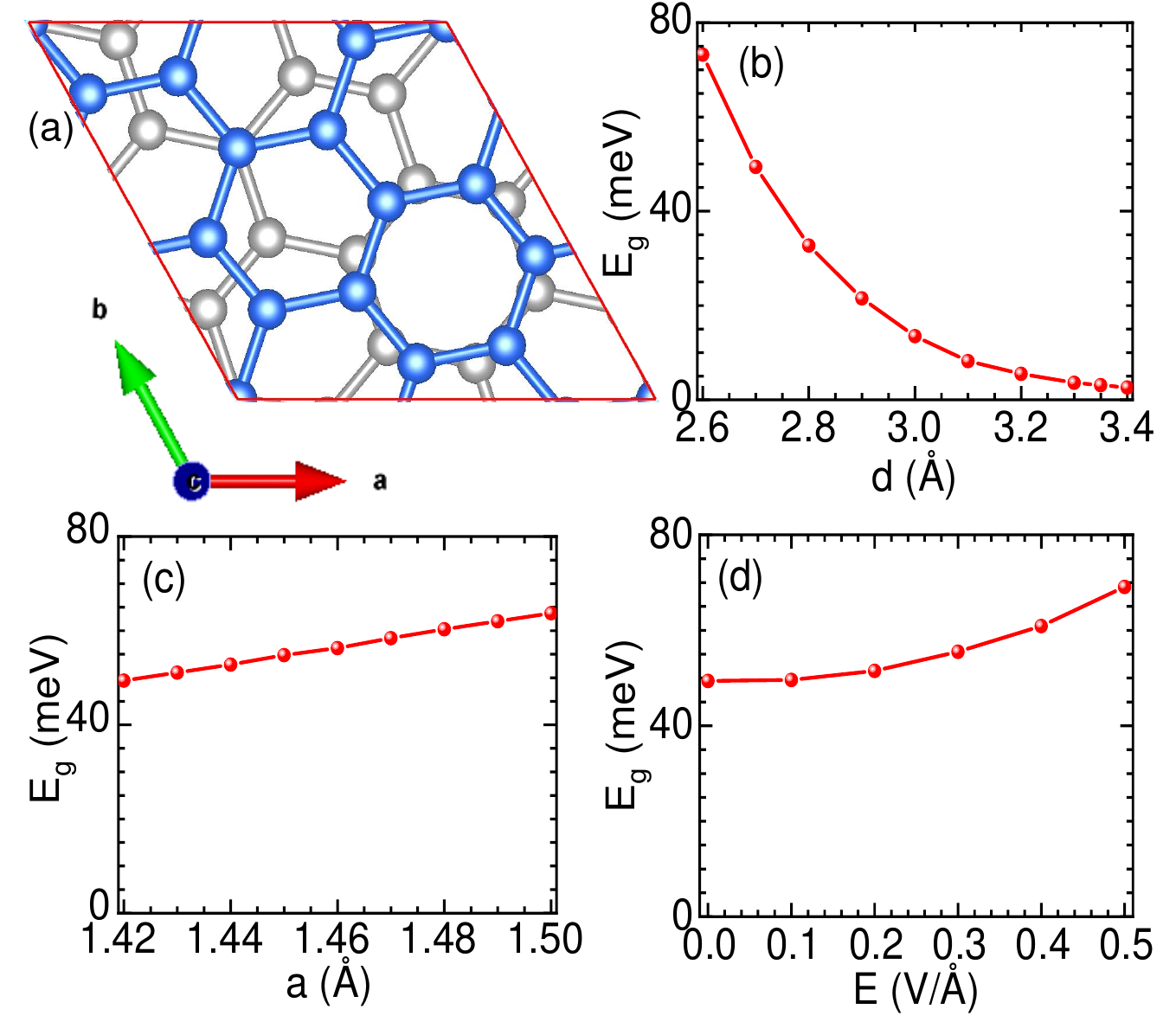}
\caption{(a) Structure of SE-even tBLG with twisted angle $\theta=38.213^{\circ}$. (b) Gap as a function of d, where a=1.42 {\AA} is used. (c) Gap as a function of a, where d=2.7 {\AA} is adopted. (d) Gap as a function of gated field $E$, where a=1.42 {\AA} and d=2.7 {\AA} are used.}
\label{fig:structure-and-gap}
\end{figure}

\begin{figure}[htbp]
\includegraphics[width=0.46\textwidth]{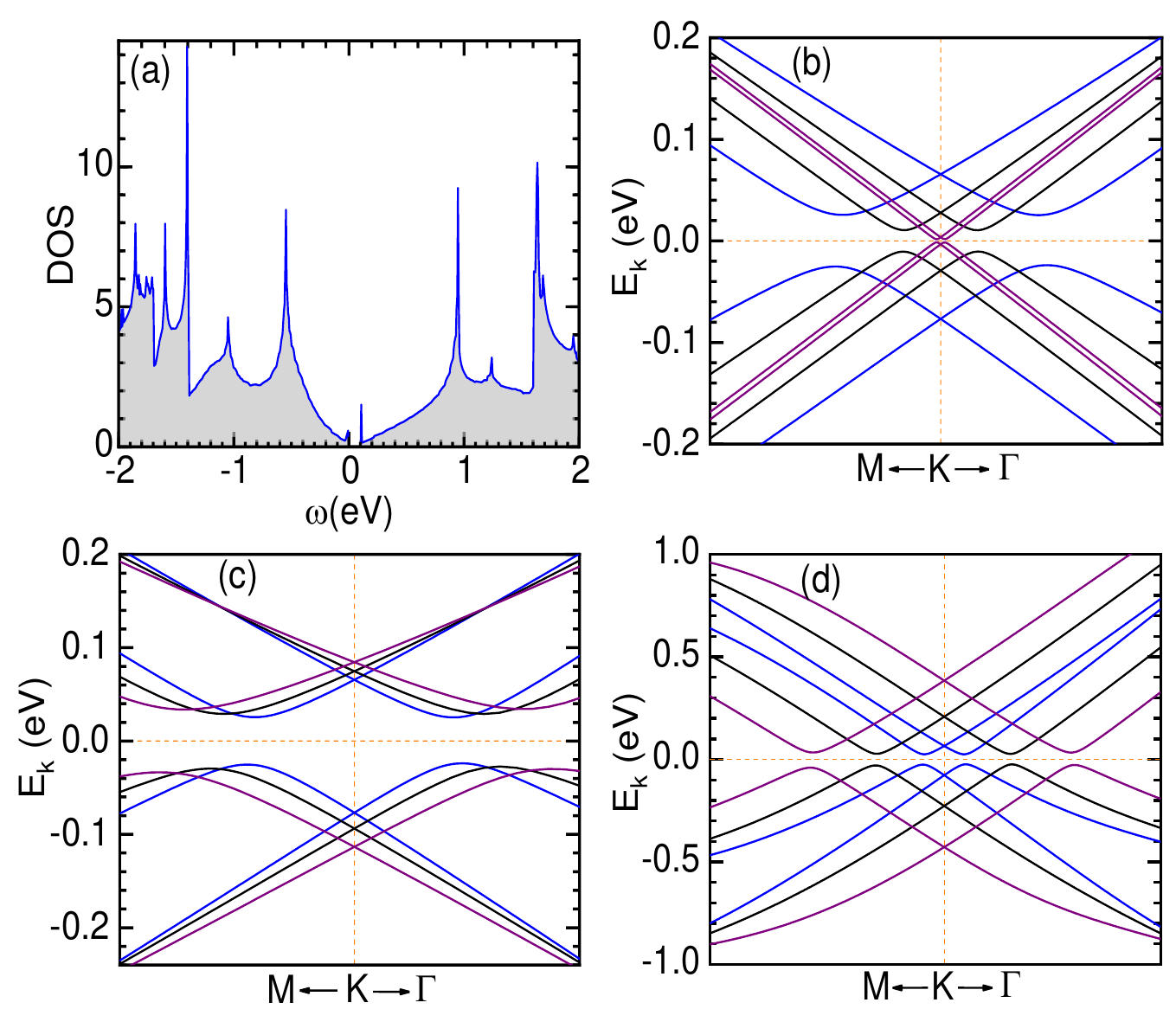}
\caption{(a) Total density of states for a=1.42 {\AA}, d=2.7 {\AA}. (b) Band structures for different inter-layer distances with 2.7 {\AA} (blue), 2.9 {\AA} (black) and 3.35 {\AA} (purple), where a=1.42 {\AA} is used. (c) Band structures for different bond lengths with 1.42 {\AA} (blue), 1.46 {\AA} (black) and 1.50 {\AA} (purple), where d=2.7 {\AA} is adopted. (d) Band structures for different gated fields  with 0.0 $ V/${\AA}(blue), 0.2 $ V/${\AA} (black) and 0.4 $ V/${\AA} (purple), where a=1.42 {\AA} and d=2.7 {\AA} are used.}
\label{BLG:bandstructure}
\end{figure}

Now, we start with the tunable properties of the gap in SE-even tBLG with twisted angle of $\theta=38.21^{\circ}$. Fig.~\ref{fig:structure-and-gap}(b) shows gap as a function of inter-layer distance $d$, where intra-layer C-C bond length $a=1.42$ {\AA} is adopted. As can be seen, the system is always gapped despite the differences in the inter-layer distance. The gap increases monotonously with decrease of $d$, which can be effectively viewed as applying transverse homogeneous pressure. We estimate the magnitude of the pressure to be 30 GPa when $d=2.7$ {\AA}. Therefore, gap of the system is tunable by a transverse homogeneous pressure. It is worth noting that gap of the equilibrium geometry of tBLG is of $3.1$~meV, which is comparable with the results obtained from other first-principles calculations~\cite{shallcross2008quantum} and long-wavelength theory~\cite{mele2010commensuration} but is not coincident with the result of tight-binding approximation where unreasonable large gap was obtained~\cite{PhysRevB.92.075402}. Band gap opening is further confirmed by DOS and band structures as presented in Fig.~\ref{BLG:bandstructure}(a) ($d=2.7$ {\AA} only) and Fig.~\ref{BLG:bandstructure}(b) respectively.

To show the effect of an in-plane biaxial tensile strain on gap opening of the system, we have calculated gap as a function of intra-layer C-C bond length $a$ as shown in Fig.~\ref{fig:structure-and-gap}(c), where $d=2.7$ {\AA} is used. We find gap increasing monotonously with the increase of $a$, indicating that gap increases monotonously with an applied in-plane biaxial tensile strain. The same conclusion is obtained by other first-principles calculations when the inter-layer distance is fixed at the equilibrium distance~\cite{khatibi2019strain}. Band gap opening is further confirmed by band structures as illustrated in Fig.~\ref{BLG:bandstructure}(c). Therefore, gap is tunable by an in-plane biaxial tensile strain.

\begin{figure}[htbp]
\includegraphics[width=0.48\textwidth]{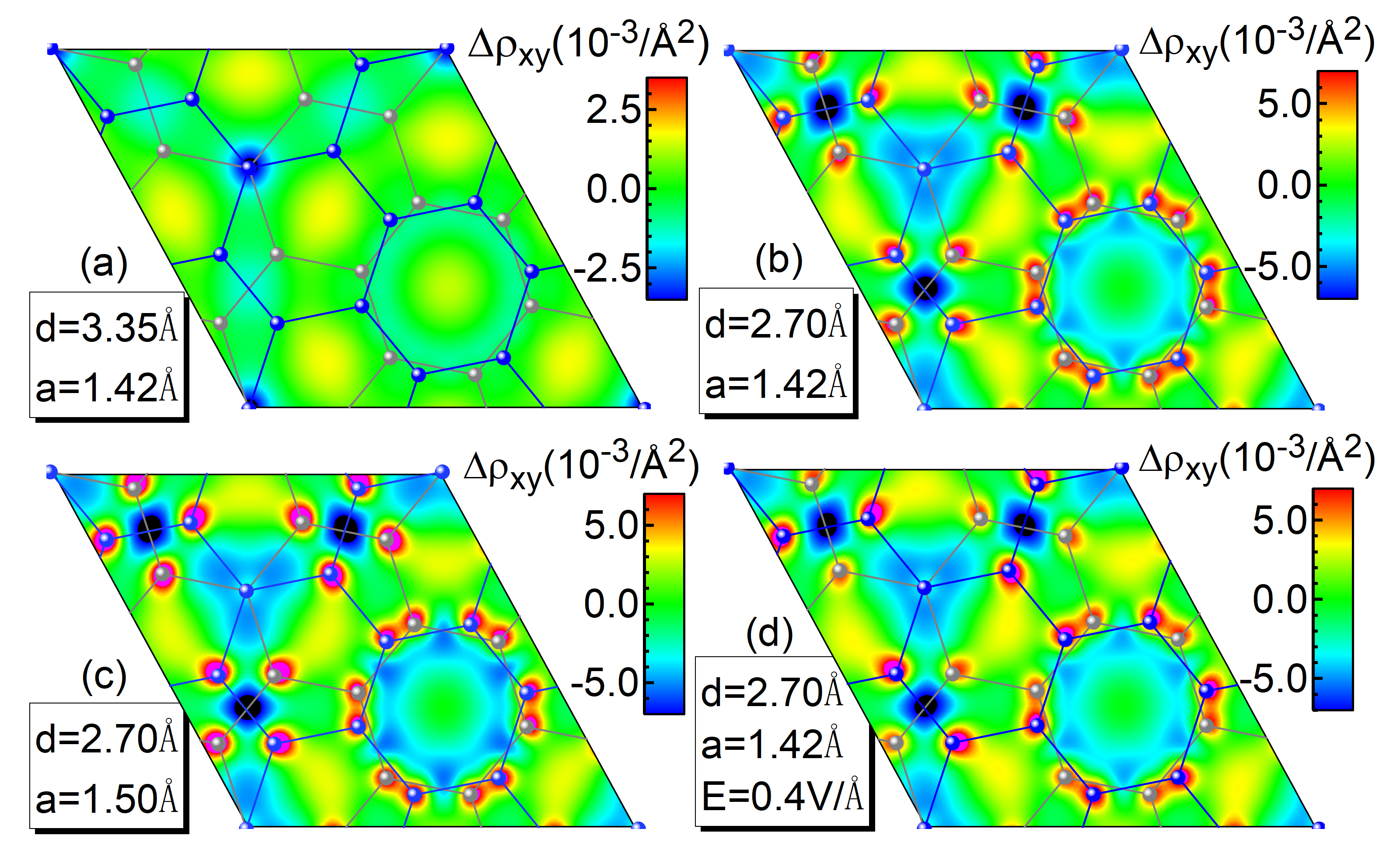}
\caption{Contour maps of charge density difference between tBLG and superposition of two individual graphene layers for different structures with d=3.35 {\AA}, a=1.42 {\AA} in (a), d=2.70 {\AA}, a=1.42 {\AA} in (b), d=2.70 {\AA}, a=1.50 {\AA} in (c) and d=2.70 {\AA}, a=1.42 {\AA}, E=0.4 $V/${\AA} in (d), respectively. Note that the scale of the color bar in (a) is half of that in (b)-(d).}
\label{fig:charge_difference}
\end{figure}
A transverse electric field can induce symmetry breaking between top and bottom layers, which may tune gap of the system. Fig.~\ref{fig:structure-and-gap}(d) shows the gap as a function of a transverse electric field, where $a=1.42$ {\AA} and $d=2.7$ {\AA} are used. We find that gap of the system increases monotonously with an applied transverse electric field, which is consistent with the results observed in experiment~\cite{liu2015observation} and that obtained by tight-binding model~\cite{sboychakov2018externally}, but is contrary to the conclusion obtained by continuum approximation where gapless state remains in the presence of gated electric field~\cite{dos2007graphene}. Band structures under different transverse electric field are presented in Fig.~\ref{BLG:bandstructure}(d). Thus, an electric field can tune the gap as well.

It is worth noting that an intrinsic intra-layer CDW exists in the system, which has not yet been mentioned. Besides, an inter-layer charge imbalance occurs when applying a transverse electric field. To illustrate this, we show the contour maps of charge density difference between tBLG and superposition of two individual graphene layers where charges are equally distributed at all sites for different structures in Fig.~\ref{fig:charge_difference}. Interestingly, there is an uneven charge distribution in tBLG even with the structure of equilibrium geometry as shown in Fig.~\ref{fig:charge_difference}(a), indicating existence of an intrinsic intra-layer CDW. As we can see from Figs.~\ref{fig:charge_difference}(b) and ~\ref{fig:charge_difference}(c), the uneven charge distribution enhances when applying a transverse homogeneous pressure and an in-plane biaxial tensile strain, respectively. Furthermore, when applying an transverse electric field, Fig.~\ref{fig:charge_difference}(d) exhibits an additional imbalance of charge distribution between top (blue) and bottom (gray) layer compared with the case without electric field (Fig.~\ref{fig:charge_difference}(b)), indicating the formation of an inter-layer charge imbalance. In fact, an intra-layer CDW or an inter-layer charge imbalance plays a crucial role in gap opening. Therefore, clarifying the effect of inter-layer hoppings, intra-layer CDW and inter-layer charge imbalance on gap opening is the key to obtain a larger gap in tBLG.

Next, we proceed to demonstrate that, the gap size increasing with reduced inter-layer distance or strengthened in-plane biaxial tensile strain is attributed to the increase of both intra-layer CDW and effective inter-layer hopping, where the latter always dominates gap opening, while gap is enlarged by the gated electric field due to the formation of an inter-layer charge imbalance.

\begin{figure}[htbp]
\includegraphics[width=0.48\textwidth,height=0.54\textwidth]{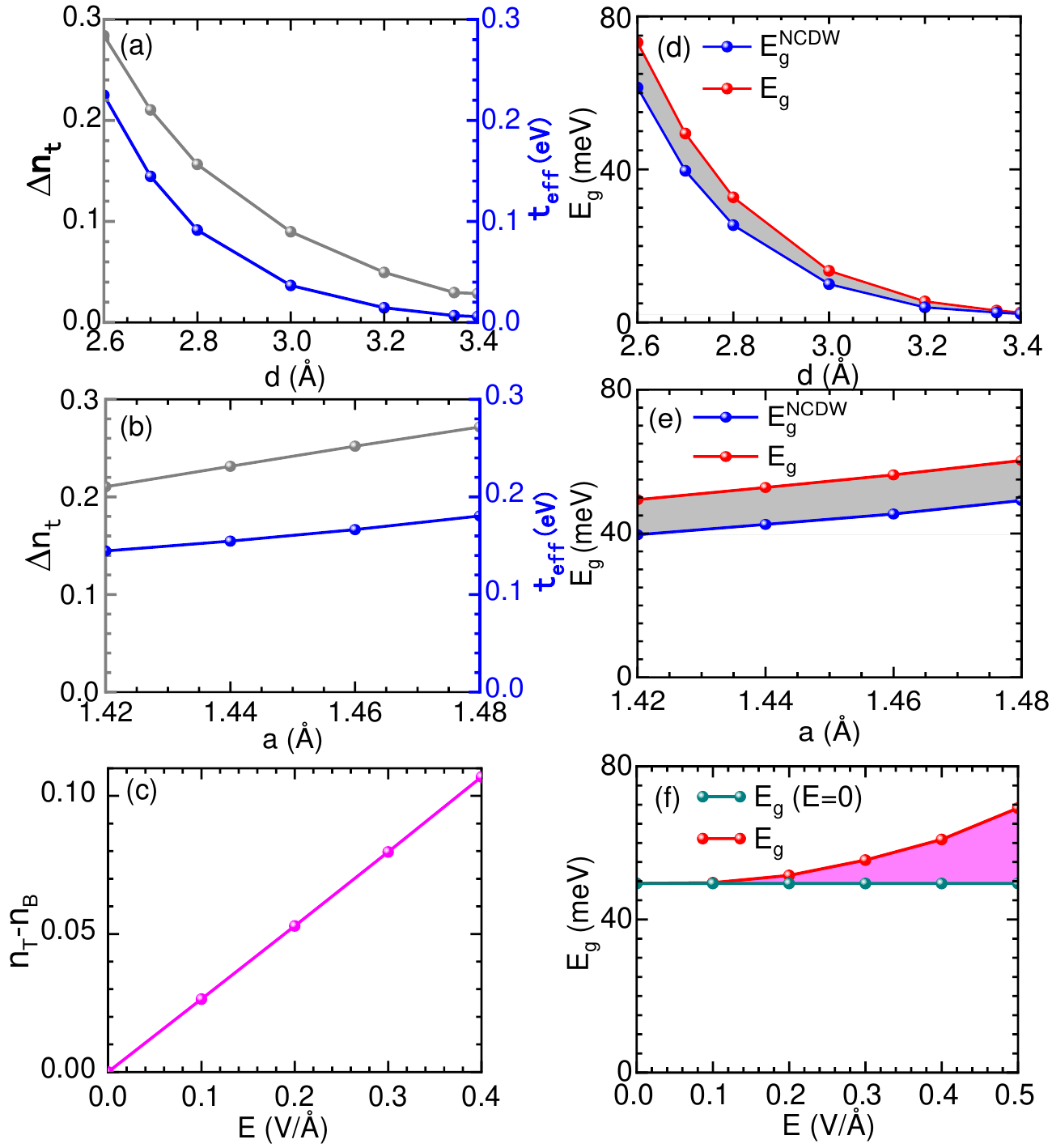}
\caption{(a) and (b) show the effective inter-layer hopping $t_{eff}$ and order parameter of the intra-layer CDW $\Delta{n_{t}}$ as functions of $d$ and $a$ respectively. (c) Order parameter of the inter-layer charge imbalance $n_{T}$-$n_B$ as a function of $E$. (d) and (e) present $E_{g}^{NCDW}$ and $E_g$ as functions of $d$ and $a$ respectively. $E_{g}^{NCDW}$ is the gap induced by inter-layer hoppings alone, where the effect of intra-layer CDW is removed artificially, while $E_g$ is the real gap of the system. NCDW is the abbreviation of no charge density wave. (f) $E_g$ as a function of $E$, where $E_g(E=0)$ is the gap without electric field. a=1.42 {\AA} is used in (a) and (d), d=2.70 {\AA} is used in (b) and (e), d=2.70 {\AA} and a=1.42 {\AA} are used in (c) and (f). The calculation method of $t_{eff}$ for different structures are shown in Fig.~\ref{fig:band_NCDW_d} of Appendix A and Fig.~\ref{fig:band_NCDW_a} of Appendix B.}
\label{fig:order-parameters}
\end{figure}

A transverse homogeneous pressure can push the top and bottom layers closer together, which enhances both inter-layer hybridizations and Coulomb interactions, leading to the increase of effective inter-layer hopping and intra-layer CDW. In order to quantify the strength of inter-layer hoppings and intra-layer CDW, we have calculated effective inter-layer hopping $t_{eff}$, defined as the band splitting at the $K$ point of Brillouin zone in the absence of intra-layer CDW, and order parameter of the intra-layer CDW $\Delta{n_{t}}=\sum_{s}|n_{s}-1.0|$ as functions of $d$ in Fig.~\ref{fig:order-parameters}(a), where $\sum_{s}$ sums over the sublattices in the supercell. Obviously, $t_{eff}$ and $\Delta{n_{t}}$ increase monotonously with reduced inter-layer distance. As a result, both the gap induced by effective inter-layer hopping alone $E_{g}^{NCDW}$, where the effect of intra-layer CDW is removed artificially, and the gap caused by intra-layer CDW $E_g-E_{g}^{NCDW}$ (gray shadow) increase with the transverse homogeneous pressure as shown in  Fig.~\ref{fig:order-parameters}(d). Thus, the real gap of the system $E_g$ increase monotonously with the transverse homogeneous pressure. Moreover, one can see that the effective inter-layer hopping dominates the gap opening.

An in-plane biaxial tensile strain stretches the in-plane C-C bond $a$, resulting in the decrease of intra-layer hybridizations, indicating that p$_z$ electrons around the C atoms are more localized in the intra-layer direction while more extended in the inter-layer direction. As a result, the effective inter-layer hopping increases due to the increased overlap between wave functions of two layers, and the intra-layer CDW is enhanced since electrons are more localized in the intra-layer direction, which are confirmed by our calculations on effective inter-layer hopping $t_{eff}$ and order parameter of the intra-layer CDW $\Delta{n_{t}}$ as functions of $a$ in Fig.~\ref{fig:order-parameters}(b). Therefore, through a similar analysis as we did in the case of applying transverse homogeneous pressure, we reveal that the $E_g$ increasing monotonously with the in-plane biaxial tensile strain, as shown in Fig.~\ref{fig:order-parameters}(e), can be ascribed to the increase of both intra-layer CDW and effective inter-layer hopping, where the latter always dominates the gap opening.
\begin{figure}[htbp]
\includegraphics[width=0.48\textwidth,height=0.22\textwidth]{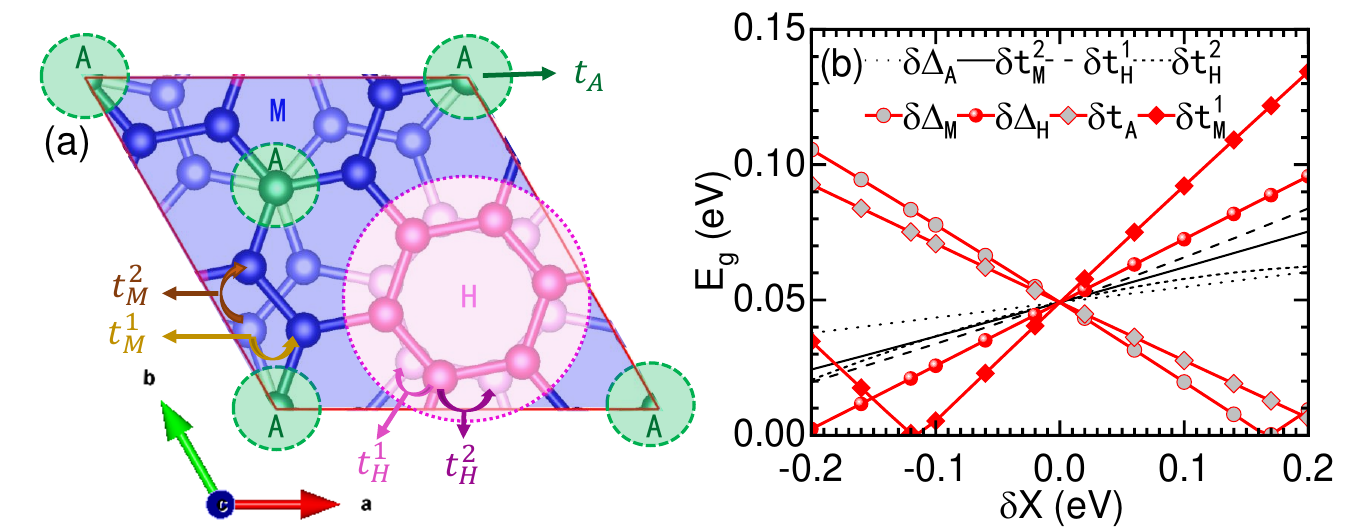}
\caption{(a)  Three distinct regions , namely, A (green) region, M (blue) region, and H (pink) region in the moi$\rm\acute{r}$e supercell. (b) Gaps as functions of the deviations of inter-layer hoppings and on-site potentials $\delta{X}$, including $\delta{t}_A$, $\delta{t}_M^1$, $\delta{t}_M^2$, $\delta{t}_H^1$, $\delta{t}_H^2$, $\delta{\Delta}_A$, $\delta{\Delta}_M$, and $\delta{\Delta}_H$ as described in Eq.~\eqref{TB_detail}, which are used to simulate the effect of external perturbations. Gap at $\delta{X}=0$ corresponds to the gap of $d=2.7$ {\AA} and $a=1.42$ {\AA}.}
\label{fig:gap-parameters}
\end{figure}

An (downward) electric field perpendicular to tBLG layers generates an electric potential difference between the top and bottom layers, which leads to electronic charge transfer from the bottom to top layer. We have calculated order parameter of inter-layer charge imbalance $n_{T}-n_B=\sum_{s}^{T}n_{s}-\sum_{s^{\prime}}^{B}n_{s^{\prime}}$ as a function of $E$ in Fig.~\ref{fig:order-parameters}(c), where $\sum_{s}^{T}$ and $\sum_{s^{\prime}}^{B}$ sum over the sublattices of top and bottom layer in the supercell, respectively. We find that $n_{T}-n_B$ increases monotonously with the electric field. As a result, the gap is further opened as shown in Fig.~\ref{fig:order-parameters}(f), where $E_g-E_g(E=0)$ (pink shadow) is the gap induced by the inter-layer charge imbalance. Therefore, we reveal that gap increasing with the electric field is due to the formation of an inter-layer charge imbalance.

In brief, gap of the system is tunable by inter-layer distance, in-plane biaxial strain and gated electric field since those can increase the effective inter-layer hopping, intra-layer CDW, or inter-layer charge imbalance. However, simply applying these external perturbations cannot open a large gap in the system. Fortunately, the effective inter-layer hopping and intra-layer CDW depend on various inter-layer hybridizations and intra-layer on-site potentials in different regions of the supercell (see Fig.~\ref{fig:gap-parameters}(a)), respectively, which provides additional degrees of freedom to tune the gap. Therefore, in the following, we will demonstrate how to open a larger gap by tuning inter-layer hybridizations and on-site potentials in specific regions of the moi$\rm\acute{r}$e supercell.

\begin{figure}[htbp]
\includegraphics[width=0.48\textwidth,height=0.46\textwidth]{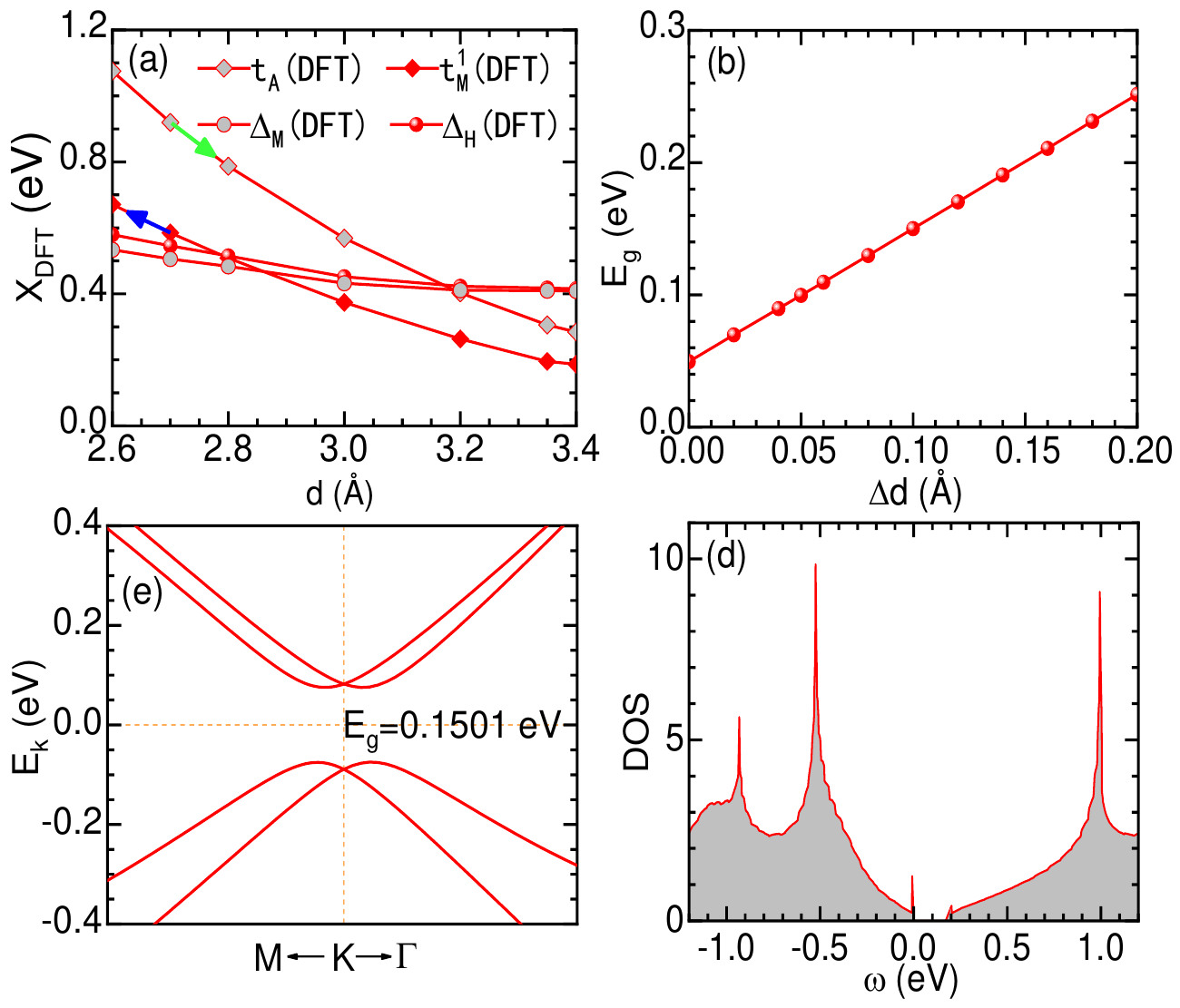}
\caption{(a) The DFT values of $t_A$, $t_M^1$, $\Delta_M$, and $\Delta_H$ as functions of $d$. (b) Gap as a function of the height difference between A and M regions within the same layer $\Delta{d}$, where inter-layer distances of $d_A=2.7$ {\AA}$+\Delta{d}$, $d_M=2.7$ {\AA}$-\Delta{d}$, and $d_H=2.7$ {\AA} are used. (c) and (d) illustrate the band structure and density of state of the case $\Delta{d}=0.1$ {\AA} respectively.}
\label{fig:large_gap}
\end{figure}

According to the lattice structure of tBLG we studied, totally five inter-layer hopping terms between neighboring sites within same region as defined in Fig.~\ref{fig:gap-parameters}(a) and three different on-site potential terms for three different regions are believed to dominate the inter-layer hoppings and intra-layer CDW. Thus, we introduce perturbations to these terms as shown in Sec.~\ref{TBLG:model_method}. In Fig.~\ref{fig:gap-parameters}(b), we present gap amplitude as functions of these perturbations at $d=2.7$ {\AA} and $a=1.42$ {\AA}. It is found that the gap increases dramatically as $\delta{t}_{M}^{1}$ and $\delta{\Delta}_H$ increase or $\delta{t}_A$ and $\delta{\Delta}_{M}$ decrease. Therefore, to open a larger gap in this system experimentally, one needs to enhance $t_{M}^{1}$ and $\Delta_H$, meanwhile weaken $t_A$ and $\Delta_H$. Since above DFT calculations demonstrate that effective inter-layer hopping dominates the gap opening, we would like to propose that applying a periodic transverse inhomogeneous pressure which affect $t_{M}^{1}$ and $t_A$ in opposite direction may be a promising way to obtain a large gap.

Finally, we will verify above proposal by first-principles calculations. In Fig.~\ref{fig:large_gap}(a), it is shown that both $t_{M}^{1}$ and $t_A$ increase with decrease of $d$, indicating that, by enlarging the inter-layer distance of A region $d_A$ and compressing that of M region $d_M$ as shown by the colored arrow in Fig.~\ref{fig:large_gap}(a), one may fulfill the requirement of increasing $t_{M}^{1}$ and decreasing $t_A$ simultaneously. To demonstrate this, we perform DFT calculations on tBLG with region-dependent inter-layer distances of $d_A=2.7${\AA}$+\Delta{d}$, $d_M=2.7${\AA}$-\Delta{d}$, and $d_H=2.7$ {\AA}, which can be effectively viewed as applying a periodic transverse homogeneous pressures. Fig.~\ref{fig:large_gap}(b) presents the gap as a function of $\Delta{d}$. It is found that the gap is larger than $100$~meV when $\Delta{d}\geq0.05$ {\AA}. Interestingly, the band structure still satisfies the nearly linear low-energy spectra as illustrated in Fig.~\ref{fig:large_gap}(c). The gap is of $150.1$~meV when $\Delta{d}=0.1$ {\AA}, which is further confirmed by the DOS in Fig.~\ref{fig:large_gap}(d). Therefore, a periodic transverse inhomogeneous pressure is an efficient way to open a larger gap in this system.

\section{DISCUSSION}\label{TBLG:DISCUSSION}

Here, combining first-principle calculations with tight-binding calculations, we point out a way to open a large gap in tBLG with twisted angle $\theta=38.21^{\circ}$. All of our calculations are performed without the consideration of spin polarization since the DOS of this system near the Fermi surface is too small to stabilize a magnetic ground state. When the twisted angle is not so large, tBLG under a large transverse pressure will favor a correlated ferromagnetic ground state due to the occurrence of the flat bands~\cite{yndurain2019pressure}. However, flat bands do not exist in our case.

Our paper mainly focuses on the tBLG with twisted angle $\theta=38.21^{\circ}$. To our knowledge, the properties of tBLG are strongly dependent on the twisted angle between two layers based on model analyses. The low-energy spectra are quite different for SE-even tBLG and SE-odd tBLG, where the former is gapped while the latter is gapless due to symmetry~\cite{PhysRevB.84.235439}. While an in-plane shift of one layer with respect to the other may turn SE-even tBLG into SE-odd tBLG resulting in gap close, SE-even tBLG is stable and can be realized in experiment~\cite{koren2016coherent}. We have also done calculations on tBLG with twisted angle $\theta=46.83^{\circ}$. It is found that the gap also increase quickly when applying a transverse homogeneous pressure as illustrated in Fig.~\ref{fig:band_46} of Appendix~\ref{app:other}. Thus, it is interesting to study the properties of tBLGs with other twisted angles based on first-principles calculations. However, it is beyond the scope of the present paper.

Interestingly, we have proposed here for the first time that there is an intrinsic intra-layer CDW in SE-even tBLG with large twisted angle, which has not yet been mentioned. In fact, the intra-layer on-site potential differences affect mainly the low-energy spectra of tBLG as shown in Fig.~\ref{fig:band_NCDW_d} of Appendix~\ref{app:efftd} and Fig.~\ref{fig:band_NCDW_a} of Appendix~\ref{app:effta}. However, the intra-layer on-site potential differences are neglected in long-wavelength theory~\cite{mele2010commensuration} and tight-binding approximation~\cite{PhysRevB.92.075402}. As a result, flat bands with broken electron-hole symmetry, which are different from that obtained by tight-binding approximation, are observed by DFT calculations~\cite{yndurain2019pressure}. Therefore, we strongly suggest that it is necessary to include the effect of intra-layer CDW in long-wavelength theory and tight-binding approximation to study the low-energy properties of tBLG.

We have shown that applying a periodic transverse inhomogeneous pressure leads to region-dependent inter-layer distances and consequently enhance the gap. Perhaps this is not difficult to achieve in experiments. It has been observed that deposition of $\rm{C_{60}}$ on tBLG can increase the gap of tBLG from 35 to 80 meV due to the increased concentration of inter-layer C-C bonding~\cite{park2015observation}. Moreover, we would like to propose that deposition of polar molecule may lead to region-dependent on-site potentials which may favor gapped state. It is worth noting that not only does the tBLG we studied can open a large band gap under a periodic transverse inhomogeneous pressure, but it still maintains nearly linear dispersion of low-energy spectra, indicating the high electron mobility still remains.

It is worth mentioning that the band gaps of semiconductors are systematically underestimated by DFT calculatons~\cite{aryasetiawan1998gw}. Thus, the actual band gap given by experiment may be larger than the calculated one in tBLG.

\section{conclusion}\label{TBLG:conclusion}
At large commensurate angles, tBLG which holds even parity under sublattice exchange exhibits a tiny gap. Here, we point out a way to tune this tiny gap into a larger gap. First, from first-principles calculations, we find that the gap size increases with increasing transverse homogeneous pressures, in-plane biaxial tensile strains, or transverse electric fields. We reveal that enhancement of the gap by pressures or strains is attributed to the increases of both intra-layer CDW and effective inter-layer hopping, while it is due to the formation of an inter-layer charge imbalance under applied electric fields. Then, using a tight-binding model, we demonstrate how to open a larger gap by tuning inter-layer hoppings and on-site potentials in specific regions of the moi$\rm\acute{r}$e supercell. Finally, based on the guidance of tight-binding calculations, we propose that a periodic transverse inhomogeneous pressure can open a gap of over 100 meV, which is further confirmed by first-principles calculations. Our results provide a theoretical guidance for experiments to open a large gap in tBLG, and may pave a way for carbon-based electronics.

\begin{center}
\textbf{ACKNOWLEDGEMENT}
\end{center}
This work is supported by National Natural Science Foundation of China (Nos.11774258, 12004283) and Shanghai Science and technology program (No.21JC405700).
\appendix
\section{The $t_{eff}$ for structures with different $d$}\label{app:efftd}
\begin{figure}[htbp]
\includegraphics[width=0.48\textwidth,height=0.48\textwidth]{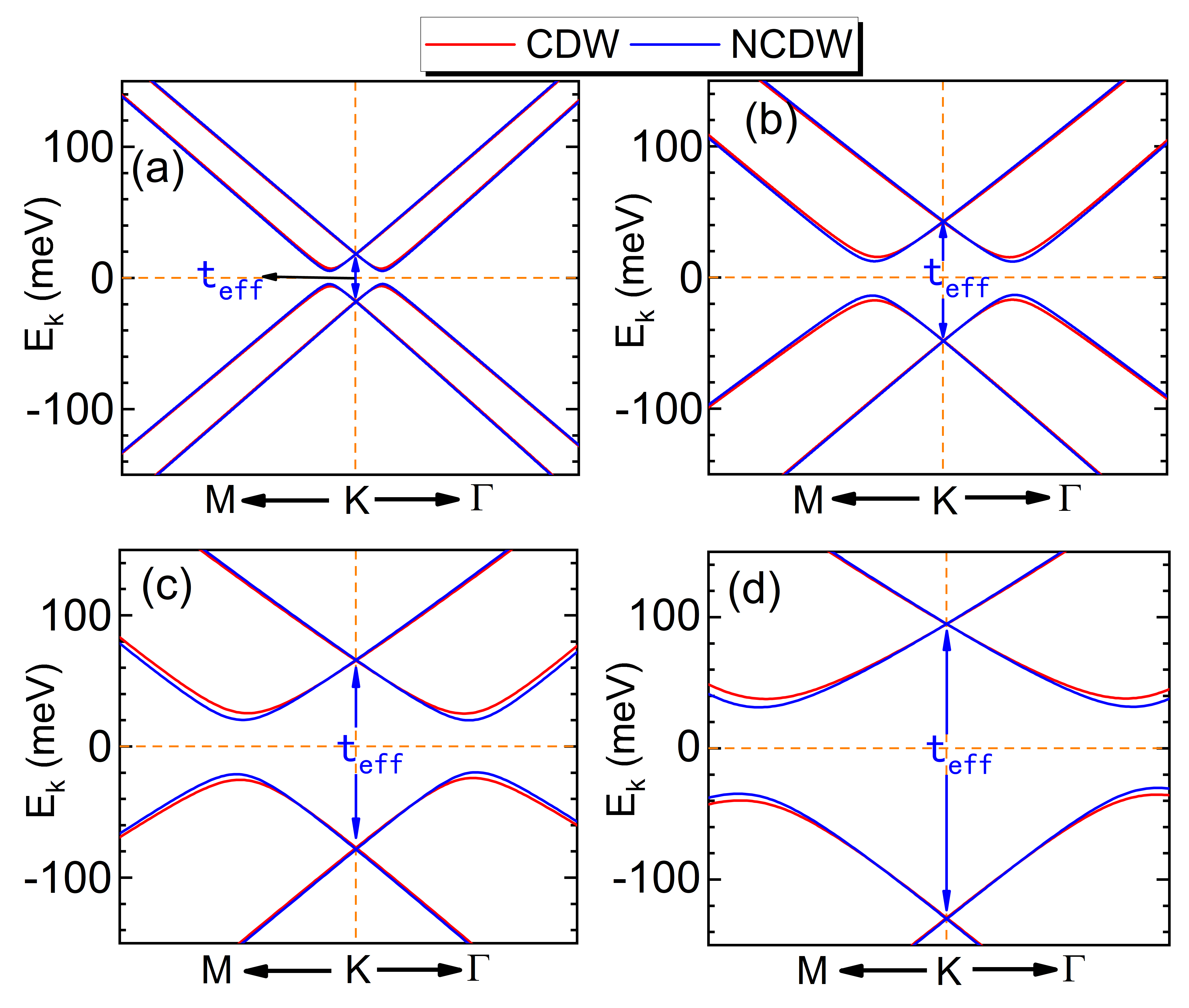}
\caption{At twisted angle $\theta=38.21^{\circ}$, band structures of tBLG with and without the effect of intra-layer CDW for different $d$, $d=3.35$ {\AA}, $d=2.8$ {\AA}, $d=2.7$ {\AA}, and $d=2.6$ {\AA} are used in (a), (b), (c), and (d) respectively. $a=1.42$ {\AA} is adopted in all cases.}
\label{fig:band_NCDW_d}
\end{figure}
Fig.~\ref{fig:band_NCDW_d} shows the band structures of tBLG with and without the effect of intra-layer CDW for structures with different $d$, where the effective inter-layer hopping $t_{eff}$ for different $d$ are given. As can be seen, $t_{eff}$ increases monotonously with the decrease of $d$.
\section{The $t_{eff}$ for structures with different $a$}\label{app:effta}
\begin{figure}[htbp]
\includegraphics[width=0.48\textwidth,height=0.48\textwidth]{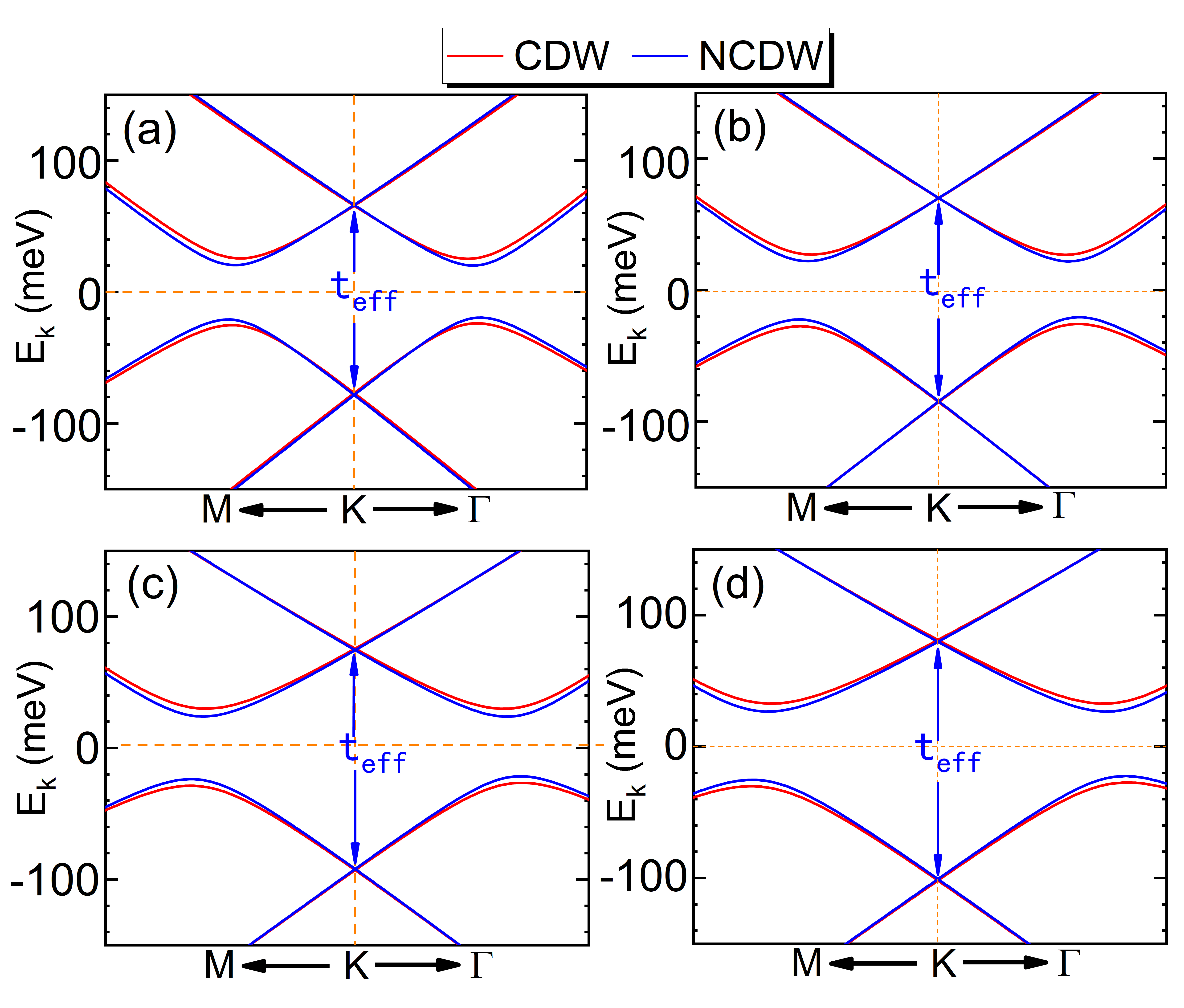}
\caption{At twisted angle $\theta=38.21^{\circ}$, band structures of tBLG with and without the effect of intra-layer CDW for different $a$, $a=1.42$ {\AA}, $a=1.44$ {\AA}, $a=1.46$ {\AA}, and $a=1.48$ {\AA} are used in (a), (b), (c), and (d) respectively. $d=2.7$ {\AA} is adopted in all cases}
\label{fig:band_NCDW_a}
\end{figure}
Fig.~\ref{fig:band_NCDW_a} shows the band structures of tBLG with and without the effect of intra-layer CDW for structures with different $a$, where the effective inter-layer hopping $t_{eff}$ for different $a$ are also given.  It is to easy find that, $t_{eff}$ increases monotonously with the increase of $a$.
\section{Tunable gap in tBLG with twisted angle $\theta=46.83^{\circ}$}\label{app:other}
\begin{figure}[htbp]
\includegraphics[width=0.48\textwidth,height=0.22\textwidth]{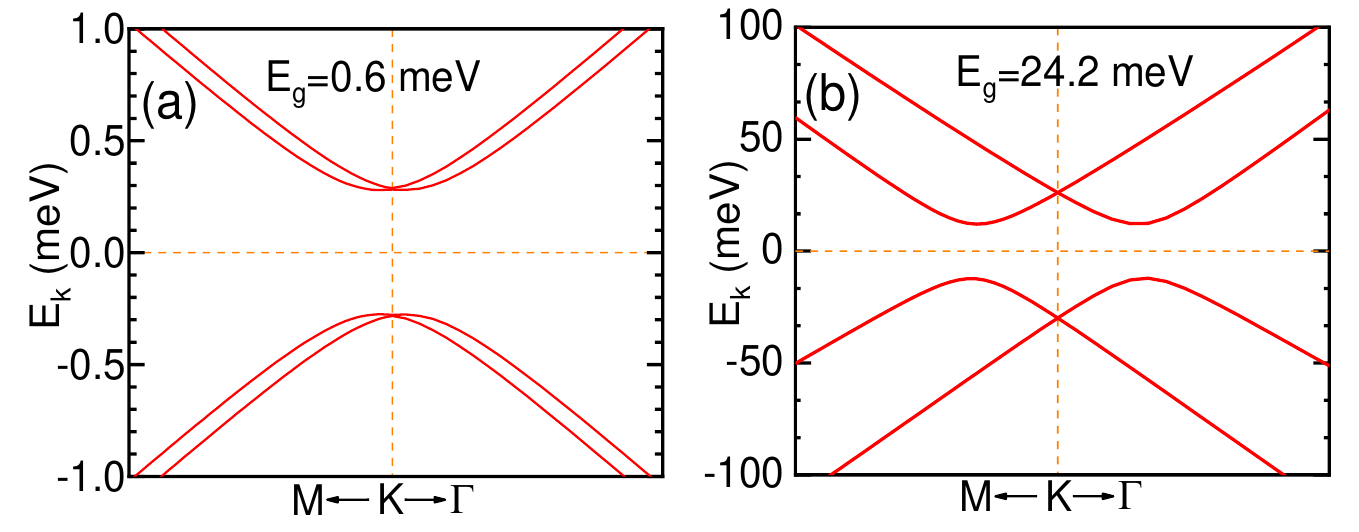}
\caption{Band structure of tBLG with twisted angle $\theta=46.83^{\circ}$, $d=3.35$ {\AA} and $d=2.7$ {\AA} are used in (a) and (b) respectively, where $a=1.42$ {\AA} is used.}
\label{fig:band_46}
\end{figure}
Fig.~\ref{fig:band_46} shows the band structures of tBLG with twisted angle $\theta=46.83^{\circ}$. When without a transverse homogeneous pressure ($d=3.35$ {\AA}), as can be seen in Fig.~\ref{fig:band_46}(a), the gap of the system $E_g=0.6$ meV. The gap increase quickly under a transverse homogeneous pressure ($d=2.7$ {\AA}), where the system is gapped with a gap of 24.2 meV in Fig.~\ref{fig:band_46}(b).
\bibliography{TBLG_reference}

\providecommand{\noopsort}[1]{}\providecommand{\singleletter}[1]{#1}%
\begin{thebibliography}{43}%
\makeatletter
\providecommand \@ifxundefined [1]{%
 \@ifx{#1\undefined}
}%
\providecommand \@ifnum [1]{%
 \ifnum #1\expandafter \@firstoftwo
 \else \expandafter \@secondoftwo
 \fi
}%
\providecommand \@ifx [1]{%
 \ifx #1\expandafter \@firstoftwo
 \else \expandafter \@secondoftwo
 \fi
}%
\providecommand \natexlab [1]{#1}%
\providecommand \enquote  [1]{``#1''}%
\providecommand \bibnamefont  [1]{#1}%
\providecommand \bibfnamefont [1]{#1}%
\providecommand \citenamefont [1]{#1}%
\providecommand \href@noop [0]{\@secondoftwo}%
\providecommand \href [0]{\begingroup \@sanitize@url \@href}%
\providecommand \@href[1]{\@@startlink{#1}\@@href}%
\providecommand \@@href[1]{\endgroup#1\@@endlink}%
\providecommand \@sanitize@url [0]{\catcode `\\12\catcode `\$12\catcode
  `\&12\catcode `\#12\catcode `\^12\catcode `\_12\catcode `\%12\relax}%
\providecommand \@@startlink[1]{}%
\providecommand \@@endlink[0]{}%
\providecommand \url  [0]{\begingroup\@sanitize@url \@url }%
\providecommand \@url [1]{\endgroup\@href {#1}{\urlprefix }}%
\providecommand \urlprefix  [0]{URL }%
\providecommand \Eprint [0]{\href }%
\providecommand \doibase [0]{http://dx.doi.org/}%
\providecommand \selectlanguage [0]{\@gobble}%
\providecommand \bibinfo  [0]{\@secondoftwo}%
\providecommand \bibfield  [0]{\@secondoftwo}%
\providecommand \translation [1]{[#1]}%
\providecommand \BibitemOpen [0]{}%
\providecommand \bibitemStop [0]{}%
\providecommand \bibitemNoStop [0]{.\EOS\space}%
\providecommand \EOS [0]{\spacefactor3000\relax}%
\providecommand \BibitemShut  [1]{\csname bibitem#1\endcsname}%
\let\auto@bib@innerbib\@empty
\bibitem [{\citenamefont {Rozhkov}\ \emph {et~al.}(2016)\citenamefont
  {Rozhkov}, \citenamefont {Sboychakov}, \citenamefont {Rakhmanov},\ and\
  \citenamefont {Nori}}]{rozhkov2016electronic}%
  \BibitemOpen
  \bibfield  {author} {\bibinfo {author} {\bibfnamefont
  {Alexandr~Vladimirovich}\ \bibnamefont {Rozhkov}}, \bibinfo {author}
  {\bibfnamefont {AO}~\bibnamefont {Sboychakov}}, \bibinfo {author}
  {\bibfnamefont {AL}~\bibnamefont {Rakhmanov}}, \ and\ \bibinfo {author}
  {\bibfnamefont {Franco}\ \bibnamefont {Nori}},\ }\bibfield  {title} {\enquote
  {\bibinfo {title} {Electronic properties of graphene-based bilayer
  systems},}\ }\href@noop {} {\bibfield  {journal} {\bibinfo  {journal}
  {Physics Reports}\ }\textbf {\bibinfo {volume} {648}},\ \bibinfo {pages}
  {1--104} (\bibinfo {year} {2016})}\BibitemShut {NoStop}%
\bibitem [{\citenamefont {Ohta}\ \emph {et~al.}(2012)\citenamefont {Ohta},
  \citenamefont {Robinson}, \citenamefont {Feibelman}, \citenamefont
  {Bostwick}, \citenamefont {Rotenberg},\ and\ \citenamefont
  {Beechem}}]{PhysRevLett.109.186807}%
  \BibitemOpen
  \bibfield  {author} {\bibinfo {author} {\bibfnamefont {Taisuke}\ \bibnamefont
  {Ohta}}, \bibinfo {author} {\bibfnamefont {Jeremy~T.}\ \bibnamefont
  {Robinson}}, \bibinfo {author} {\bibfnamefont {Peter~J.}\ \bibnamefont
  {Feibelman}}, \bibinfo {author} {\bibfnamefont {Aaron}\ \bibnamefont
  {Bostwick}}, \bibinfo {author} {\bibfnamefont {Eli}\ \bibnamefont
  {Rotenberg}}, \ and\ \bibinfo {author} {\bibfnamefont {Thomas~E.}\
  \bibnamefont {Beechem}},\ }\bibfield  {title} {\enquote {\bibinfo {title}
  {Evidence for interlayer coupling and moir\'e periodic potentials in twisted
  bilayer graphene},}\ }\href {\doibase 10.1103/PhysRevLett.109.186807}
  {\bibfield  {journal} {\bibinfo  {journal} {Phys. Rev. Lett.}\ }\textbf
  {\bibinfo {volume} {109}},\ \bibinfo {pages} {186807} (\bibinfo {year}
  {2012})}\BibitemShut {NoStop}%
\bibitem [{\citenamefont {Cao}\ \emph {et~al.}(2018{\natexlab{a}})\citenamefont
  {Cao}, \citenamefont {Fatemi}, \citenamefont {Demir}, \citenamefont {Fang},
  \citenamefont {Tomarken}, \citenamefont {Luo}, \citenamefont
  {Sanchez-Yamagishi}, \citenamefont {Watanabe}, \citenamefont {Taniguchi},
  \citenamefont {Kaxiras} \emph {et~al.}}]{cao2018correlated}%
  \BibitemOpen
  \bibfield  {author} {\bibinfo {author} {\bibfnamefont {Yuan}\ \bibnamefont
  {Cao}}, \bibinfo {author} {\bibfnamefont {Valla}\ \bibnamefont {Fatemi}},
  \bibinfo {author} {\bibfnamefont {Ahmet}\ \bibnamefont {Demir}}, \bibinfo
  {author} {\bibfnamefont {Shiang}\ \bibnamefont {Fang}}, \bibinfo {author}
  {\bibfnamefont {Spencer~L}\ \bibnamefont {Tomarken}}, \bibinfo {author}
  {\bibfnamefont {Jason~Y}\ \bibnamefont {Luo}}, \bibinfo {author}
  {\bibfnamefont {Javier~D}\ \bibnamefont {Sanchez-Yamagishi}}, \bibinfo
  {author} {\bibfnamefont {Kenji}\ \bibnamefont {Watanabe}}, \bibinfo {author}
  {\bibfnamefont {Takashi}\ \bibnamefont {Taniguchi}}, \bibinfo {author}
  {\bibfnamefont {Efthimios}\ \bibnamefont {Kaxiras}},  \emph {et~al.},\
  }\bibfield  {title} {\enquote {\bibinfo {title} {Correlated insulator
  behaviour at half-filling in magic-angle graphene superlattices},}\
  }\href@noop {} {\bibfield  {journal} {\bibinfo  {journal} {Nature}\ }\textbf
  {\bibinfo {volume} {556}},\ \bibinfo {pages} {80--84} (\bibinfo {year}
  {2018}{\natexlab{a}})}\BibitemShut {NoStop}%
\bibitem [{\citenamefont {Po}\ \emph {et~al.}(2018)\citenamefont {Po},
  \citenamefont {Zou}, \citenamefont {Vishwanath},\ and\ \citenamefont
  {Senthil}}]{po2018origin}%
  \BibitemOpen
  \bibfield  {author} {\bibinfo {author} {\bibfnamefont {Hoi~Chun}\
  \bibnamefont {Po}}, \bibinfo {author} {\bibfnamefont {Liujun}\ \bibnamefont
  {Zou}}, \bibinfo {author} {\bibfnamefont {Ashvin}\ \bibnamefont
  {Vishwanath}}, \ and\ \bibinfo {author} {\bibfnamefont {T}~\bibnamefont
  {Senthil}},\ }\bibfield  {title} {\enquote {\bibinfo {title} {Origin of mott
  insulating behavior and superconductivity in twisted bilayer graphene},}\
  }\href@noop {} {\bibfield  {journal} {\bibinfo  {journal} {Physical Review
  X}\ }\textbf {\bibinfo {volume} {8}},\ \bibinfo {pages} {031089} (\bibinfo
  {year} {2018})}\BibitemShut {NoStop}%
\bibitem [{\citenamefont {Cao}\ \emph {et~al.}(2018{\natexlab{b}})\citenamefont
  {Cao}, \citenamefont {Fatemi}, \citenamefont {Fang}, \citenamefont
  {Watanabe}, \citenamefont {Taniguchi}, \citenamefont {Kaxiras},\ and\
  \citenamefont {Jarillo-Herrero}}]{cao2018unconventional}%
  \BibitemOpen
  \bibfield  {author} {\bibinfo {author} {\bibfnamefont {Yuan}\ \bibnamefont
  {Cao}}, \bibinfo {author} {\bibfnamefont {Valla}\ \bibnamefont {Fatemi}},
  \bibinfo {author} {\bibfnamefont {Shiang}\ \bibnamefont {Fang}}, \bibinfo
  {author} {\bibfnamefont {Kenji}\ \bibnamefont {Watanabe}}, \bibinfo {author}
  {\bibfnamefont {Takashi}\ \bibnamefont {Taniguchi}}, \bibinfo {author}
  {\bibfnamefont {Efthimios}\ \bibnamefont {Kaxiras}}, \ and\ \bibinfo {author}
  {\bibfnamefont {Pablo}\ \bibnamefont {Jarillo-Herrero}},\ }\bibfield  {title}
  {\enquote {\bibinfo {title} {Unconventional superconductivity in magic-angle
  graphene superlattices},}\ }\href@noop {} {\bibfield  {journal} {\bibinfo
  {journal} {Nature}\ }\textbf {\bibinfo {volume} {556}},\ \bibinfo {pages}
  {43--50} (\bibinfo {year} {2018}{\natexlab{b}})}\BibitemShut {NoStop}%
\bibitem [{\citenamefont {Lu}\ \emph {et~al.}(2019)\citenamefont {Lu},
  \citenamefont {Stepanov}, \citenamefont {Yang}, \citenamefont {Xie},
  \citenamefont {Aamir}, \citenamefont {Das}, \citenamefont {Urgell},
  \citenamefont {Watanabe}, \citenamefont {Taniguchi}, \citenamefont {Zhang}
  \emph {et~al.}}]{lu2019superconductors}%
  \BibitemOpen
  \bibfield  {author} {\bibinfo {author} {\bibfnamefont {Xiaobo}\ \bibnamefont
  {Lu}}, \bibinfo {author} {\bibfnamefont {Petr}\ \bibnamefont {Stepanov}},
  \bibinfo {author} {\bibfnamefont {Wei}\ \bibnamefont {Yang}}, \bibinfo
  {author} {\bibfnamefont {Ming}\ \bibnamefont {Xie}}, \bibinfo {author}
  {\bibfnamefont {Mohammed~Ali}\ \bibnamefont {Aamir}}, \bibinfo {author}
  {\bibfnamefont {Ipsita}\ \bibnamefont {Das}}, \bibinfo {author}
  {\bibfnamefont {Carles}\ \bibnamefont {Urgell}}, \bibinfo {author}
  {\bibfnamefont {Kenji}\ \bibnamefont {Watanabe}}, \bibinfo {author}
  {\bibfnamefont {Takashi}\ \bibnamefont {Taniguchi}}, \bibinfo {author}
  {\bibfnamefont {Guangyu}\ \bibnamefont {Zhang}},  \emph {et~al.},\ }\bibfield
   {title} {\enquote {\bibinfo {title} {Superconductors, orbital magnets and
  correlated states in magic-angle bilayer graphene},}\ }\href@noop {}
  {\bibfield  {journal} {\bibinfo  {journal} {Nature}\ }\textbf {\bibinfo
  {volume} {574}},\ \bibinfo {pages} {653--657} (\bibinfo {year}
  {2019})}\BibitemShut {NoStop}%
\bibitem [{\citenamefont {Balents}\ \emph {et~al.}(2020)\citenamefont
  {Balents}, \citenamefont {Dean}, \citenamefont {Efetov},\ and\ \citenamefont
  {Young}}]{balents2020superconductivity}%
  \BibitemOpen
  \bibfield  {author} {\bibinfo {author} {\bibfnamefont {Leon}\ \bibnamefont
  {Balents}}, \bibinfo {author} {\bibfnamefont {Cory~R}\ \bibnamefont {Dean}},
  \bibinfo {author} {\bibfnamefont {Dmitri~K}\ \bibnamefont {Efetov}}, \ and\
  \bibinfo {author} {\bibfnamefont {Andrea~F}\ \bibnamefont {Young}},\
  }\bibfield  {title} {\enquote {\bibinfo {title} {Superconductivity and strong
  correlations in moir{\'e} flat bands},}\ }\href@noop {} {\bibfield  {journal}
  {\bibinfo  {journal} {Nature Physics}\ }\textbf {\bibinfo {volume} {16}},\
  \bibinfo {pages} {725--733} (\bibinfo {year} {2020})}\BibitemShut {NoStop}%
\bibitem [{\citenamefont {Park}\ \emph {et~al.}(2019)\citenamefont {Park},
  \citenamefont {Kim}, \citenamefont {Cho},\ and\ \citenamefont
  {Lee}}]{park2019higher}%
  \BibitemOpen
  \bibfield  {author} {\bibinfo {author} {\bibfnamefont {Moon~Jip}\
  \bibnamefont {Park}}, \bibinfo {author} {\bibfnamefont {Youngkuk}\
  \bibnamefont {Kim}}, \bibinfo {author} {\bibfnamefont {Gil~Young}\
  \bibnamefont {Cho}}, \ and\ \bibinfo {author} {\bibfnamefont {SungBin}\
  \bibnamefont {Lee}},\ }\bibfield  {title} {\enquote {\bibinfo {title}
  {Higher-order topological insulator in twisted bilayer graphene},}\
  }\href@noop {} {\bibfield  {journal} {\bibinfo  {journal} {Physical review
  letters}\ }\textbf {\bibinfo {volume} {123}},\ \bibinfo {pages} {216803}
  (\bibinfo {year} {2019})}\BibitemShut {NoStop}%
\bibitem [{\citenamefont {Kindermann}(2015)}]{kindermann2015topological}%
  \BibitemOpen
  \bibfield  {author} {\bibinfo {author} {\bibfnamefont {M}~\bibnamefont
  {Kindermann}},\ }\bibfield  {title} {\enquote {\bibinfo {title} {Topological
  crystalline insulator phase in graphene multilayers},}\ }\href@noop {}
  {\bibfield  {journal} {\bibinfo  {journal} {Physical review letters}\
  }\textbf {\bibinfo {volume} {114}},\ \bibinfo {pages} {226802} (\bibinfo
  {year} {2015})}\BibitemShut {NoStop}%
\bibitem [{\citenamefont {Sprinkle}\ \emph {et~al.}(2009)\citenamefont
  {Sprinkle}, \citenamefont {Siegel}, \citenamefont {Hu}, \citenamefont
  {Hicks}, \citenamefont {Tejeda}, \citenamefont {Taleb-Ibrahimi},
  \citenamefont {Le~F{\`e}vre}, \citenamefont {Bertran}, \citenamefont
  {Vizzini}, \citenamefont {Enriquez} \emph {et~al.}}]{sprinkle2009first}%
  \BibitemOpen
  \bibfield  {author} {\bibinfo {author} {\bibfnamefont {Mike}\ \bibnamefont
  {Sprinkle}}, \bibinfo {author} {\bibfnamefont {David}\ \bibnamefont
  {Siegel}}, \bibinfo {author} {\bibfnamefont {Yike}\ \bibnamefont {Hu}},
  \bibinfo {author} {\bibfnamefont {J}~\bibnamefont {Hicks}}, \bibinfo {author}
  {\bibfnamefont {Antonio}\ \bibnamefont {Tejeda}}, \bibinfo {author}
  {\bibfnamefont {Amina}\ \bibnamefont {Taleb-Ibrahimi}}, \bibinfo {author}
  {\bibfnamefont {Patrick}\ \bibnamefont {Le~F{\`e}vre}}, \bibinfo {author}
  {\bibfnamefont {Fran{\c{c}}ois}\ \bibnamefont {Bertran}}, \bibinfo {author}
  {\bibfnamefont {S}~\bibnamefont {Vizzini}}, \bibinfo {author} {\bibfnamefont
  {H}~\bibnamefont {Enriquez}},  \emph {et~al.},\ }\bibfield  {title} {\enquote
  {\bibinfo {title} {First direct observation of a nearly ideal graphene band
  structure},}\ }\href@noop {} {\bibfield  {journal} {\bibinfo  {journal}
  {Physical Review Letters}\ }\textbf {\bibinfo {volume} {103}},\ \bibinfo
  {pages} {226803} (\bibinfo {year} {2009})}\BibitemShut {NoStop}%
\bibitem [{\citenamefont {Shallcross}\ \emph {et~al.}(2008)\citenamefont
  {Shallcross}, \citenamefont {Sharma},\ and\ \citenamefont
  {Pankratov}}]{shallcross2008quantum}%
  \BibitemOpen
  \bibfield  {author} {\bibinfo {author} {\bibfnamefont {S}~\bibnamefont
  {Shallcross}}, \bibinfo {author} {\bibfnamefont {Sangeeta}\ \bibnamefont
  {Sharma}}, \ and\ \bibinfo {author} {\bibfnamefont {Oleg~A}\ \bibnamefont
  {Pankratov}},\ }\bibfield  {title} {\enquote {\bibinfo {title} {Quantum
  interference at the twist boundary in graphene},}\ }\href@noop {} {\bibfield
  {journal} {\bibinfo  {journal} {Physical review letters}\ }\textbf {\bibinfo
  {volume} {101}},\ \bibinfo {pages} {056803} (\bibinfo {year}
  {2008})}\BibitemShut {NoStop}%
\bibitem [{\citenamefont {Shallcross}\ \emph {et~al.}(2010)\citenamefont
  {Shallcross}, \citenamefont {Sharma}, \citenamefont {Kandelaki},\ and\
  \citenamefont {Pankratov}}]{PhysRevB.81.165105}%
  \BibitemOpen
  \bibfield  {author} {\bibinfo {author} {\bibfnamefont {S.}~\bibnamefont
  {Shallcross}}, \bibinfo {author} {\bibfnamefont {S.}~\bibnamefont {Sharma}},
  \bibinfo {author} {\bibfnamefont {E.}~\bibnamefont {Kandelaki}}, \ and\
  \bibinfo {author} {\bibfnamefont {O.~A.}\ \bibnamefont {Pankratov}},\
  }\bibfield  {title} {\enquote {\bibinfo {title} {Electronic structure of
  turbostratic graphene},}\ }\href {\doibase 10.1103/PhysRevB.81.165105}
  {\bibfield  {journal} {\bibinfo  {journal} {Phys. Rev. B}\ }\textbf {\bibinfo
  {volume} {81}},\ \bibinfo {pages} {165105} (\bibinfo {year}
  {2010})}\BibitemShut {NoStop}%
\bibitem [{\citenamefont {Oostinga}\ \emph {et~al.}(2008)\citenamefont
  {Oostinga}, \citenamefont {Heersche}, \citenamefont {Liu}, \citenamefont
  {Morpurgo},\ and\ \citenamefont {Vandersypen}}]{oostinga2008gate}%
  \BibitemOpen
  \bibfield  {author} {\bibinfo {author} {\bibfnamefont {Jeroen~B}\
  \bibnamefont {Oostinga}}, \bibinfo {author} {\bibfnamefont {Hubert~B}\
  \bibnamefont {Heersche}}, \bibinfo {author} {\bibfnamefont {Xinglan}\
  \bibnamefont {Liu}}, \bibinfo {author} {\bibfnamefont {Alberto~F}\
  \bibnamefont {Morpurgo}}, \ and\ \bibinfo {author} {\bibfnamefont
  {Lieven~MK}\ \bibnamefont {Vandersypen}},\ }\bibfield  {title} {\enquote
  {\bibinfo {title} {Gate-induced insulating state in bilayer graphene
  devices},}\ }\href@noop {} {\bibfield  {journal} {\bibinfo  {journal} {Nature
  materials}\ }\textbf {\bibinfo {volume} {7}},\ \bibinfo {pages} {151--157}
  (\bibinfo {year} {2008})}\BibitemShut {NoStop}%
\bibitem [{\citenamefont {Mele}(2010)}]{mele2010commensuration}%
  \BibitemOpen
  \bibfield  {author} {\bibinfo {author} {\bibfnamefont {Eugene~J}\
  \bibnamefont {Mele}},\ }\bibfield  {title} {\enquote {\bibinfo {title}
  {Commensuration and interlayer coherence in twisted bilayer graphene},}\
  }\href@noop {} {\bibfield  {journal} {\bibinfo  {journal} {Physical Review
  B}\ }\textbf {\bibinfo {volume} {81}},\ \bibinfo {pages} {161405} (\bibinfo
  {year} {2010})}\BibitemShut {NoStop}%
\bibitem [{\citenamefont {Mele}(2012)}]{mele2012interlayer}%
  \BibitemOpen
  \bibfield  {author} {\bibinfo {author} {\bibfnamefont {EJ}~\bibnamefont
  {Mele}},\ }\bibfield  {title} {\enquote {\bibinfo {title} {Interlayer
  coupling in rotationally faulted multilayer graphenes},}\ }\href@noop {}
  {\bibfield  {journal} {\bibinfo  {journal} {Journal of Physics D: Applied
  Physics}\ }\textbf {\bibinfo {volume} {45}},\ \bibinfo {pages} {154004}
  (\bibinfo {year} {2012})}\BibitemShut {NoStop}%
\bibitem [{\citenamefont {Sboychakov}\ \emph {et~al.}(2015)\citenamefont
  {Sboychakov}, \citenamefont {Rakhmanov}, \citenamefont {Rozhkov},\ and\
  \citenamefont {Nori}}]{PhysRevB.92.075402}%
  \BibitemOpen
  \bibfield  {author} {\bibinfo {author} {\bibfnamefont {A.~O.}\ \bibnamefont
  {Sboychakov}}, \bibinfo {author} {\bibfnamefont {A.~L.}\ \bibnamefont
  {Rakhmanov}}, \bibinfo {author} {\bibfnamefont {A.~V.}\ \bibnamefont
  {Rozhkov}}, \ and\ \bibinfo {author} {\bibfnamefont {Franco}\ \bibnamefont
  {Nori}},\ }\bibfield  {title} {\enquote {\bibinfo {title} {Electronic
  spectrum of twisted bilayer graphene},}\ }\href {\doibase
  10.1103/PhysRevB.92.075402} {\bibfield  {journal} {\bibinfo  {journal} {Phys.
  Rev. B}\ }\textbf {\bibinfo {volume} {92}},\ \bibinfo {pages} {075402}
  (\bibinfo {year} {2015})}\BibitemShut {NoStop}%
\bibitem [{\citenamefont {Khatibi}\ \emph {et~al.}(2019)\citenamefont
  {Khatibi}, \citenamefont {Namiranian},\ and\ \citenamefont
  {Parhizgar}}]{khatibi2019strain}%
  \BibitemOpen
  \bibfield  {author} {\bibinfo {author} {\bibfnamefont {Zahra}\ \bibnamefont
  {Khatibi}}, \bibinfo {author} {\bibfnamefont {Afshin}\ \bibnamefont
  {Namiranian}}, \ and\ \bibinfo {author} {\bibfnamefont {Fariborz}\
  \bibnamefont {Parhizgar}},\ }\bibfield  {title} {\enquote {\bibinfo {title}
  {Strain impacts on commensurate bilayer graphene superlattices: Distorted
  trigonal warping, emergence of bandgap and direct-indirect bandgap
  transition},}\ }\href@noop {} {\bibfield  {journal} {\bibinfo  {journal}
  {Diamond and Related Materials}\ }\textbf {\bibinfo {volume} {92}},\ \bibinfo
  {pages} {228--234} (\bibinfo {year} {2019})}\BibitemShut {NoStop}%
\bibitem [{\citenamefont {Rozhkov}\ \emph {et~al.}(2017)\citenamefont
  {Rozhkov}, \citenamefont {Sboychakov}, \citenamefont {Rakhmanov},\ and\
  \citenamefont {Nori}}]{rozhkov2017single}%
  \BibitemOpen
  \bibfield  {author} {\bibinfo {author} {\bibfnamefont {AV}~\bibnamefont
  {Rozhkov}}, \bibinfo {author} {\bibfnamefont {AO}~\bibnamefont {Sboychakov}},
  \bibinfo {author} {\bibfnamefont {AL}~\bibnamefont {Rakhmanov}}, \ and\
  \bibinfo {author} {\bibfnamefont {Franco}\ \bibnamefont {Nori}},\ }\bibfield
  {title} {\enquote {\bibinfo {title} {Single-electron gap in the spectrum of
  twisted bilayer graphene},}\ }\href@noop {} {\bibfield  {journal} {\bibinfo
  {journal} {Physical Review B}\ }\textbf {\bibinfo {volume} {95}},\ \bibinfo
  {pages} {045119} (\bibinfo {year} {2017})}\BibitemShut {NoStop}%
\bibitem [{\citenamefont {Cao}\ \emph {et~al.}(2016)\citenamefont {Cao},
  \citenamefont {Luo}, \citenamefont {Fatemi}, \citenamefont {Fang},
  \citenamefont {Sanchez-Yamagishi}, \citenamefont {Watanabe}, \citenamefont
  {Taniguchi}, \citenamefont {Kaxiras},\ and\ \citenamefont
  {Jarillo-Herrero}}]{cao2016superlattice}%
  \BibitemOpen
  \bibfield  {author} {\bibinfo {author} {\bibfnamefont {Y}~\bibnamefont
  {Cao}}, \bibinfo {author} {\bibfnamefont {JY}~\bibnamefont {Luo}}, \bibinfo
  {author} {\bibfnamefont {V}~\bibnamefont {Fatemi}}, \bibinfo {author}
  {\bibfnamefont {S}~\bibnamefont {Fang}}, \bibinfo {author} {\bibfnamefont
  {JD}~\bibnamefont {Sanchez-Yamagishi}}, \bibinfo {author} {\bibfnamefont
  {K}~\bibnamefont {Watanabe}}, \bibinfo {author} {\bibfnamefont
  {T}~\bibnamefont {Taniguchi}}, \bibinfo {author} {\bibfnamefont
  {E}~\bibnamefont {Kaxiras}}, \ and\ \bibinfo {author} {\bibfnamefont {Pablo}\
  \bibnamefont {Jarillo-Herrero}},\ }\bibfield  {title} {\enquote {\bibinfo
  {title} {Superlattice-induced insulating states and valley-protected orbits
  in twisted bilayer graphene},}\ }\href@noop {} {\bibfield  {journal}
  {\bibinfo  {journal} {Physical review letters}\ }\textbf {\bibinfo {volume}
  {117}},\ \bibinfo {pages} {116804} (\bibinfo {year} {2016})}\BibitemShut
  {NoStop}%
\bibitem [{\citenamefont {Kim}\ \emph {et~al.}(2016)\citenamefont {Kim},
  \citenamefont {Yankowitz}, \citenamefont {Fallahazad}, \citenamefont {Kang},
  \citenamefont {Movva}, \citenamefont {Huang}, \citenamefont {Larentis},
  \citenamefont {Corbet}, \citenamefont {Taniguchi}, \citenamefont {Watanabe}
  \emph {et~al.}}]{kim2016van}%
  \BibitemOpen
  \bibfield  {author} {\bibinfo {author} {\bibfnamefont {Kyounghwan}\
  \bibnamefont {Kim}}, \bibinfo {author} {\bibfnamefont {Matthew}\ \bibnamefont
  {Yankowitz}}, \bibinfo {author} {\bibfnamefont {Babak}\ \bibnamefont
  {Fallahazad}}, \bibinfo {author} {\bibfnamefont {Sangwoo}\ \bibnamefont
  {Kang}}, \bibinfo {author} {\bibfnamefont {Hema~CP}\ \bibnamefont {Movva}},
  \bibinfo {author} {\bibfnamefont {Shengqiang}\ \bibnamefont {Huang}},
  \bibinfo {author} {\bibfnamefont {Stefano}\ \bibnamefont {Larentis}},
  \bibinfo {author} {\bibfnamefont {Chris~M}\ \bibnamefont {Corbet}}, \bibinfo
  {author} {\bibfnamefont {Takashi}\ \bibnamefont {Taniguchi}}, \bibinfo
  {author} {\bibfnamefont {Kenji}\ \bibnamefont {Watanabe}},  \emph {et~al.},\
  }\bibfield  {title} {\enquote {\bibinfo {title} {van der waals
  heterostructures with high accuracy rotational alignment},}\ }\href@noop {}
  {\bibfield  {journal} {\bibinfo  {journal} {Nano letters}\ }\textbf {\bibinfo
  {volume} {16}},\ \bibinfo {pages} {1989--1995} (\bibinfo {year}
  {2016})}\BibitemShut {NoStop}%
\bibitem [{\citenamefont {Kim}\ \emph {et~al.}(2017)\citenamefont {Kim},
  \citenamefont {DaSilva}, \citenamefont {Huang}, \citenamefont {Fallahazad},
  \citenamefont {Larentis}, \citenamefont {Taniguchi}, \citenamefont
  {Watanabe}, \citenamefont {LeRoy}, \citenamefont {MacDonald},\ and\
  \citenamefont {Tutuc}}]{kim2017tunable}%
  \BibitemOpen
  \bibfield  {author} {\bibinfo {author} {\bibfnamefont {Kyounghwan}\
  \bibnamefont {Kim}}, \bibinfo {author} {\bibfnamefont {Ashley}\ \bibnamefont
  {DaSilva}}, \bibinfo {author} {\bibfnamefont {Shengqiang}\ \bibnamefont
  {Huang}}, \bibinfo {author} {\bibfnamefont {Babak}\ \bibnamefont
  {Fallahazad}}, \bibinfo {author} {\bibfnamefont {Stefano}\ \bibnamefont
  {Larentis}}, \bibinfo {author} {\bibfnamefont {Takashi}\ \bibnamefont
  {Taniguchi}}, \bibinfo {author} {\bibfnamefont {Kenji}\ \bibnamefont
  {Watanabe}}, \bibinfo {author} {\bibfnamefont {Brian~J}\ \bibnamefont
  {LeRoy}}, \bibinfo {author} {\bibfnamefont {Allan~H}\ \bibnamefont
  {MacDonald}}, \ and\ \bibinfo {author} {\bibfnamefont {Emanuel}\ \bibnamefont
  {Tutuc}},\ }\bibfield  {title} {\enquote {\bibinfo {title} {Tunable moir{\'e}
  bands and strong correlations in small-twist-angle bilayer graphene},}\
  }\href@noop {} {\bibfield  {journal} {\bibinfo  {journal} {Proceedings of the
  National Academy of Sciences}\ }\textbf {\bibinfo {volume} {114}},\ \bibinfo
  {pages} {3364--3369} (\bibinfo {year} {2017})}\BibitemShut {NoStop}%
\bibitem [{\citenamefont {Zhao}\ and\ \citenamefont {Spain}(1989)}]{zhao1989x}%
  \BibitemOpen
  \bibfield  {author} {\bibinfo {author} {\bibfnamefont {You~Xiang}\
  \bibnamefont {Zhao}}\ and\ \bibinfo {author} {\bibfnamefont {Ian~L}\
  \bibnamefont {Spain}},\ }\bibfield  {title} {\enquote {\bibinfo {title}
  {X-ray diffraction data for graphite to 20 gpa},}\ }\href@noop {} {\bibfield
  {journal} {\bibinfo  {journal} {Physical Review B}\ }\textbf {\bibinfo
  {volume} {40}},\ \bibinfo {pages} {993} (\bibinfo {year} {1989})}\BibitemShut
  {NoStop}%
\bibitem [{\citenamefont {Cao}\ \emph {et~al.}(2020)\citenamefont {Cao},
  \citenamefont {Feng}, \citenamefont {Han}, \citenamefont {Gao}, \citenamefont
  {Ly}, \citenamefont {Xu},\ and\ \citenamefont {Lu}}]{cao2020elastic}%
  \BibitemOpen
  \bibfield  {author} {\bibinfo {author} {\bibfnamefont {Ke}~\bibnamefont
  {Cao}}, \bibinfo {author} {\bibfnamefont {Shizhe}\ \bibnamefont {Feng}},
  \bibinfo {author} {\bibfnamefont {Ying}\ \bibnamefont {Han}}, \bibinfo
  {author} {\bibfnamefont {Libo}\ \bibnamefont {Gao}}, \bibinfo {author}
  {\bibfnamefont {Thuc~Hue}\ \bibnamefont {Ly}}, \bibinfo {author}
  {\bibfnamefont {Zhiping}\ \bibnamefont {Xu}}, \ and\ \bibinfo {author}
  {\bibfnamefont {Yang}\ \bibnamefont {Lu}},\ }\bibfield  {title} {\enquote
  {\bibinfo {title} {Elastic straining of free-standing monolayer graphene},}\
  }\href@noop {} {\bibfield  {journal} {\bibinfo  {journal} {Nature
  communications}\ }\textbf {\bibinfo {volume} {11}},\ \bibinfo {pages} {1--7}
  (\bibinfo {year} {2020})}\BibitemShut {NoStop}%
\bibitem [{\citenamefont {Liu}\ \emph {et~al.}(2015)\citenamefont {Liu},
  \citenamefont {Li}, \citenamefont {Chen}, \citenamefont {Wang}, \citenamefont
  {Qi}, \citenamefont {He}, \citenamefont {Zheng}, \citenamefont {Zhou},
  \citenamefont {Zhang}, \citenamefont {Gu} \emph
  {et~al.}}]{liu2015observation}%
  \BibitemOpen
  \bibfield  {author} {\bibinfo {author} {\bibfnamefont {Jing-Bo}\ \bibnamefont
  {Liu}}, \bibinfo {author} {\bibfnamefont {Ping-Jian}\ \bibnamefont {Li}},
  \bibinfo {author} {\bibfnamefont {Yuan-Fu}\ \bibnamefont {Chen}}, \bibinfo
  {author} {\bibfnamefont {Ze-Gao}\ \bibnamefont {Wang}}, \bibinfo {author}
  {\bibfnamefont {Fei}\ \bibnamefont {Qi}}, \bibinfo {author} {\bibfnamefont
  {Jia-Rui}\ \bibnamefont {He}}, \bibinfo {author} {\bibfnamefont {Bin-Jie}\
  \bibnamefont {Zheng}}, \bibinfo {author} {\bibfnamefont {Jin-Hao}\
  \bibnamefont {Zhou}}, \bibinfo {author} {\bibfnamefont {Wan-Li}\ \bibnamefont
  {Zhang}}, \bibinfo {author} {\bibfnamefont {Lin}\ \bibnamefont {Gu}},  \emph
  {et~al.},\ }\bibfield  {title} {\enquote {\bibinfo {title} {Observation of
  tunable electrical bandgap in large-area twisted bilayer graphene synthesized
  by chemical vapor deposition},}\ }\href@noop {} {\bibfield  {journal}
  {\bibinfo  {journal} {Scientific reports}\ }\textbf {\bibinfo {volume} {5}},\
  \bibinfo {pages} {1--9} (\bibinfo {year} {2015})}\BibitemShut {NoStop}%
\bibitem [{\citenamefont {Carr}\ \emph {et~al.}(2018)\citenamefont {Carr},
  \citenamefont {Fang}, \citenamefont {Jarillo-Herrero},\ and\ \citenamefont
  {Kaxiras}}]{carr2018pressure}%
  \BibitemOpen
  \bibfield  {author} {\bibinfo {author} {\bibfnamefont {Stephen}\ \bibnamefont
  {Carr}}, \bibinfo {author} {\bibfnamefont {Shiang}\ \bibnamefont {Fang}},
  \bibinfo {author} {\bibfnamefont {Pablo}\ \bibnamefont {Jarillo-Herrero}}, \
  and\ \bibinfo {author} {\bibfnamefont {Efthimios}\ \bibnamefont {Kaxiras}},\
  }\bibfield  {title} {\enquote {\bibinfo {title} {Pressure dependence of the
  magic twist angle in graphene superlattices},}\ }\href@noop {} {\bibfield
  {journal} {\bibinfo  {journal} {Physical Review B}\ }\textbf {\bibinfo
  {volume} {98}},\ \bibinfo {pages} {085144} (\bibinfo {year}
  {2018})}\BibitemShut {NoStop}%
\bibitem [{\citenamefont {Ge}\ \emph {et~al.}(2021)\citenamefont {Ge},
  \citenamefont {Ni}, \citenamefont {Wu}, \citenamefont {Fu}, \citenamefont
  {Lu},\ and\ \citenamefont {Zhu}}]{ge2021emerging}%
  \BibitemOpen
  \bibfield  {author} {\bibinfo {author} {\bibfnamefont {Liangbing}\
  \bibnamefont {Ge}}, \bibinfo {author} {\bibfnamefont {Kun}\ \bibnamefont
  {Ni}}, \bibinfo {author} {\bibfnamefont {Xiaojun}\ \bibnamefont {Wu}},
  \bibinfo {author} {\bibfnamefont {Zhengping}\ \bibnamefont {Fu}}, \bibinfo
  {author} {\bibfnamefont {Yalin}\ \bibnamefont {Lu}}, \ and\ \bibinfo {author}
  {\bibfnamefont {Yanwu}\ \bibnamefont {Zhu}},\ }\bibfield  {title} {\enquote
  {\bibinfo {title} {Emerging flat bands in large-angle twisted bi-layer
  graphene under pressure},}\ }\href@noop {} {\bibfield  {journal} {\bibinfo
  {journal} {Nanoscale}\ }\textbf {\bibinfo {volume} {13}},\ \bibinfo {pages}
  {9264--9269} (\bibinfo {year} {2021})}\BibitemShut {NoStop}%
\bibitem [{\citenamefont {Yndurain}(2019)}]{yndurain2019pressure}%
  \BibitemOpen
  \bibfield  {author} {\bibinfo {author} {\bibfnamefont {Felix}\ \bibnamefont
  {Yndurain}},\ }\bibfield  {title} {\enquote {\bibinfo {title}
  {Pressure-induced magnetism in rotated graphene bilayers},}\ }\href@noop {}
  {\bibfield  {journal} {\bibinfo  {journal} {Physical Review B}\ }\textbf
  {\bibinfo {volume} {99}},\ \bibinfo {pages} {045423} (\bibinfo {year}
  {2019})}\BibitemShut {NoStop}%
\bibitem [{\citenamefont {Tepliakov}\ \emph {et~al.}(2021)\citenamefont
  {Tepliakov}, \citenamefont {Wu},\ and\ \citenamefont
  {Yazyev}}]{tepliakov2021crystal}%
  \BibitemOpen
  \bibfield  {author} {\bibinfo {author} {\bibfnamefont {Nikita~V}\
  \bibnamefont {Tepliakov}}, \bibinfo {author} {\bibfnamefont {QuanSheng}\
  \bibnamefont {Wu}}, \ and\ \bibinfo {author} {\bibfnamefont {Oleg~V}\
  \bibnamefont {Yazyev}},\ }\bibfield  {title} {\enquote {\bibinfo {title}
  {Crystal field effect and electric field screening in multilayer graphene
  with and without twist},}\ }\href@noop {} {\bibfield  {journal} {\bibinfo
  {journal} {Nano Letters}\ } (\bibinfo {year} {2021})}\BibitemShut {NoStop}%
\bibitem [{\citenamefont {Koren}\ \emph {et~al.}(2016)\citenamefont {Koren},
  \citenamefont {Leven}, \citenamefont {L{\"o}rtscher}, \citenamefont {Knoll},
  \citenamefont {Hod},\ and\ \citenamefont {Duerig}}]{koren2016coherent}%
  \BibitemOpen
  \bibfield  {author} {\bibinfo {author} {\bibfnamefont {Elad}\ \bibnamefont
  {Koren}}, \bibinfo {author} {\bibfnamefont {Itai}\ \bibnamefont {Leven}},
  \bibinfo {author} {\bibfnamefont {Emanuel}\ \bibnamefont {L{\"o}rtscher}},
  \bibinfo {author} {\bibfnamefont {Armin}\ \bibnamefont {Knoll}}, \bibinfo
  {author} {\bibfnamefont {Oded}\ \bibnamefont {Hod}}, \ and\ \bibinfo {author}
  {\bibfnamefont {Urs}\ \bibnamefont {Duerig}},\ }\bibfield  {title} {\enquote
  {\bibinfo {title} {Coherent commensurate electronic states at the interface
  between misoriented graphene layers},}\ }\href@noop {} {\bibfield  {journal}
  {\bibinfo  {journal} {Nature nanotechnology}\ }\textbf {\bibinfo {volume}
  {11}},\ \bibinfo {pages} {752--757} (\bibinfo {year} {2016})}\BibitemShut
  {NoStop}%
\bibitem [{\citenamefont {Bl\"ochl}(1994)}]{PhysRevB.50.17953}%
  \BibitemOpen
  \bibfield  {author} {\bibinfo {author} {\bibfnamefont {P.~E.}\ \bibnamefont
  {Bl\"ochl}},\ }\bibfield  {title} {\enquote {\bibinfo {title} {Projector
  augmented-wave method},}\ }\href {\doibase 10.1103/PhysRevB.50.17953}
  {\bibfield  {journal} {\bibinfo  {journal} {Phys. Rev. B}\ }\textbf {\bibinfo
  {volume} {50}},\ \bibinfo {pages} {17953--17979} (\bibinfo {year}
  {1994})}\BibitemShut {NoStop}%
\bibitem [{\citenamefont {Kresse}\ and\ \citenamefont
  {Furthm\"uller}(1996)}]{PhysRevB.54.11169}%
  \BibitemOpen
  \bibfield  {author} {\bibinfo {author} {\bibfnamefont {G.}~\bibnamefont
  {Kresse}}\ and\ \bibinfo {author} {\bibfnamefont {J.}~\bibnamefont
  {Furthm\"uller}},\ }\bibfield  {title} {\enquote {\bibinfo {title} {Efficient
  iterative schemes for ab initio total-energy calculations using a plane-wave
  basis set},}\ }\href {\doibase 10.1103/PhysRevB.54.11169} {\bibfield
  {journal} {\bibinfo  {journal} {Phys. Rev. B}\ }\textbf {\bibinfo {volume}
  {54}},\ \bibinfo {pages} {11169--11186} (\bibinfo {year} {1996})}\BibitemShut
  {NoStop}%
\bibitem [{\citenamefont {Kresse}\ and\ \citenamefont
  {Furthm{\"u}ller}(1996)}]{kresse1996efficiency}%
  \BibitemOpen
  \bibfield  {author} {\bibinfo {author} {\bibfnamefont {Georg}\ \bibnamefont
  {Kresse}}\ and\ \bibinfo {author} {\bibfnamefont {J{\"u}rgen}\ \bibnamefont
  {Furthm{\"u}ller}},\ }\bibfield  {title} {\enquote {\bibinfo {title}
  {Efficiency of ab-initio total energy calculations for metals and
  semiconductors using a plane-wave basis set},}\ }\href@noop {} {\bibfield
  {journal} {\bibinfo  {journal} {Computational materials science}\ }\textbf
  {\bibinfo {volume} {6}},\ \bibinfo {pages} {15--50} (\bibinfo {year}
  {1996})}\BibitemShut {NoStop}%
\bibitem [{\citenamefont {Perdew}\ \emph {et~al.}(1996)\citenamefont {Perdew},
  \citenamefont {Burke},\ and\ \citenamefont
  {Ernzerhof}}]{PhysRevLett.77.3865}%
  \BibitemOpen
  \bibfield  {author} {\bibinfo {author} {\bibfnamefont {John~P.}\ \bibnamefont
  {Perdew}}, \bibinfo {author} {\bibfnamefont {Kieron}\ \bibnamefont {Burke}},
  \ and\ \bibinfo {author} {\bibfnamefont {Matthias}\ \bibnamefont
  {Ernzerhof}},\ }\bibfield  {title} {\enquote {\bibinfo {title} {Generalized
  gradient approximation made simple},}\ }\href {\doibase
  10.1103/PhysRevLett.77.3865} {\bibfield  {journal} {\bibinfo  {journal}
  {Phys. Rev. Lett.}\ }\textbf {\bibinfo {volume} {77}},\ \bibinfo {pages}
  {3865--3868} (\bibinfo {year} {1996})}\BibitemShut {NoStop}%
\bibitem [{\citenamefont {Lee}\ \emph {et~al.}(2010)\citenamefont {Lee},
  \citenamefont {Murray}, \citenamefont {Kong}, \citenamefont {Lundqvist},\
  and\ \citenamefont {Langreth}}]{PhysRevB.82.081101}%
  \BibitemOpen
  \bibfield  {author} {\bibinfo {author} {\bibfnamefont {Kyuho}\ \bibnamefont
  {Lee}}, \bibinfo {author} {\bibfnamefont {\'Eamonn~D.}\ \bibnamefont
  {Murray}}, \bibinfo {author} {\bibfnamefont {Lingzhu}\ \bibnamefont {Kong}},
  \bibinfo {author} {\bibfnamefont {Bengt~I.}\ \bibnamefont {Lundqvist}}, \
  and\ \bibinfo {author} {\bibfnamefont {David~C.}\ \bibnamefont {Langreth}},\
  }\bibfield  {title} {\enquote {\bibinfo {title} {Higher-accuracy van der
  waals density functional},}\ }\href {\doibase 10.1103/PhysRevB.82.081101}
  {\bibfield  {journal} {\bibinfo  {journal} {Phys. Rev. B}\ }\textbf {\bibinfo
  {volume} {82}},\ \bibinfo {pages} {081101} (\bibinfo {year}
  {2010})}\BibitemShut {NoStop}%
\bibitem [{\citenamefont {Hamada}(2014)}]{PhysRevB.89.121103}%
  \BibitemOpen
  \bibfield  {author} {\bibinfo {author} {\bibfnamefont {Ikutaro}\ \bibnamefont
  {Hamada}},\ }\bibfield  {title} {\enquote {\bibinfo {title} {van der waals
  density functional made accurate},}\ }\href {\doibase
  10.1103/PhysRevB.89.121103} {\bibfield  {journal} {\bibinfo  {journal} {Phys.
  Rev. B}\ }\textbf {\bibinfo {volume} {89}},\ \bibinfo {pages} {121103}
  (\bibinfo {year} {2014})}\BibitemShut {NoStop}%
\bibitem [{\citenamefont {Del~Grande}\ \emph {et~al.}(2019)\citenamefont
  {Del~Grande}, \citenamefont {Menezes},\ and\ \citenamefont
  {Capaz}}]{del2019layer}%
  \BibitemOpen
  \bibfield  {author} {\bibinfo {author} {\bibfnamefont {Rafael~R}\
  \bibnamefont {Del~Grande}}, \bibinfo {author} {\bibfnamefont {Marcos~G}\
  \bibnamefont {Menezes}}, \ and\ \bibinfo {author} {\bibfnamefont {Rodrigo~B}\
  \bibnamefont {Capaz}},\ }\bibfield  {title} {\enquote {\bibinfo {title}
  {Layer breathing and shear modes in multilayer graphene: a dft-vdw study},}\
  }\href@noop {} {\bibfield  {journal} {\bibinfo  {journal} {Journal of
  Physics: Condensed Matter}\ }\textbf {\bibinfo {volume} {31}},\ \bibinfo
  {pages} {295301} (\bibinfo {year} {2019})}\BibitemShut {NoStop}%
\bibitem [{\citenamefont {Marzari}\ \emph {et~al.}(2012)\citenamefont
  {Marzari}, \citenamefont {Mostofi}, \citenamefont {Yates}, \citenamefont
  {Souza},\ and\ \citenamefont {Vanderbilt}}]{RevModPhys.84.1419}%
  \BibitemOpen
  \bibfield  {author} {\bibinfo {author} {\bibfnamefont {Nicola}\ \bibnamefont
  {Marzari}}, \bibinfo {author} {\bibfnamefont {Arash~A.}\ \bibnamefont
  {Mostofi}}, \bibinfo {author} {\bibfnamefont {Jonathan~R.}\ \bibnamefont
  {Yates}}, \bibinfo {author} {\bibfnamefont {Ivo}\ \bibnamefont {Souza}}, \
  and\ \bibinfo {author} {\bibfnamefont {David}\ \bibnamefont {Vanderbilt}},\
  }\bibfield  {title} {\enquote {\bibinfo {title} {Maximally localized wannier
  functions: Theory and applications},}\ }\href {\doibase
  10.1103/RevModPhys.84.1419} {\bibfield  {journal} {\bibinfo  {journal} {Rev.
  Mod. Phys.}\ }\textbf {\bibinfo {volume} {84}},\ \bibinfo {pages}
  {1419--1475} (\bibinfo {year} {2012})}\BibitemShut {NoStop}%
\bibitem [{\citenamefont {Mostofi}\ \emph {et~al.}(2008)\citenamefont
  {Mostofi}, \citenamefont {Yates}, \citenamefont {Lee}, \citenamefont {Souza},
  \citenamefont {Vanderbilt},\ and\ \citenamefont
  {Marzari}}]{mostofi2008wannier90}%
  \BibitemOpen
  \bibfield  {author} {\bibinfo {author} {\bibfnamefont {Arash~A}\ \bibnamefont
  {Mostofi}}, \bibinfo {author} {\bibfnamefont {Jonathan~R}\ \bibnamefont
  {Yates}}, \bibinfo {author} {\bibfnamefont {Young-Su}\ \bibnamefont {Lee}},
  \bibinfo {author} {\bibfnamefont {Ivo}\ \bibnamefont {Souza}}, \bibinfo
  {author} {\bibfnamefont {David}\ \bibnamefont {Vanderbilt}}, \ and\ \bibinfo
  {author} {\bibfnamefont {Nicola}\ \bibnamefont {Marzari}},\ }\bibfield
  {title} {\enquote {\bibinfo {title} {wannier90: A tool for obtaining
  maximally-localised wannier functions},}\ }\href@noop {} {\bibfield
  {journal} {\bibinfo  {journal} {Computer physics communications}\ }\textbf
  {\bibinfo {volume} {178}},\ \bibinfo {pages} {685--699} (\bibinfo {year}
  {2008})}\BibitemShut {NoStop}%
\bibitem [{\citenamefont {Sboychakov}\ \emph {et~al.}(2018)\citenamefont
  {Sboychakov}, \citenamefont {Rozhkov}, \citenamefont {Rakhmanov},\ and\
  \citenamefont {Nori}}]{sboychakov2018externally}%
  \BibitemOpen
  \bibfield  {author} {\bibinfo {author} {\bibfnamefont {AO}~\bibnamefont
  {Sboychakov}}, \bibinfo {author} {\bibfnamefont {AV}~\bibnamefont {Rozhkov}},
  \bibinfo {author} {\bibfnamefont {AL}~\bibnamefont {Rakhmanov}}, \ and\
  \bibinfo {author} {\bibfnamefont {Franco}\ \bibnamefont {Nori}},\ }\bibfield
  {title} {\enquote {\bibinfo {title} {Externally controlled magnetism and band
  gap in twisted bilayer graphene},}\ }\href@noop {} {\bibfield  {journal}
  {\bibinfo  {journal} {Physical review letters}\ }\textbf {\bibinfo {volume}
  {120}},\ \bibinfo {pages} {266402} (\bibinfo {year} {2018})}\BibitemShut
  {NoStop}%
\bibitem [{\citenamefont {Dos~Santos}\ \emph {et~al.}(2007)\citenamefont
  {Dos~Santos}, \citenamefont {Peres},\ and\ \citenamefont
  {Neto}}]{dos2007graphene}%
  \BibitemOpen
  \bibfield  {author} {\bibinfo {author} {\bibfnamefont {JMB~Lopes}\
  \bibnamefont {Dos~Santos}}, \bibinfo {author} {\bibfnamefont {NMR}\
  \bibnamefont {Peres}}, \ and\ \bibinfo {author} {\bibfnamefont {AH~Castro}\
  \bibnamefont {Neto}},\ }\bibfield  {title} {\enquote {\bibinfo {title}
  {Graphene bilayer with a twist: Electronic structure},}\ }\href@noop {}
  {\bibfield  {journal} {\bibinfo  {journal} {Physical review letters}\
  }\textbf {\bibinfo {volume} {99}},\ \bibinfo {pages} {256802} (\bibinfo
  {year} {2007})}\BibitemShut {NoStop}%
\bibitem [{\citenamefont {Mele}(2011)}]{PhysRevB.84.235439}%
  \BibitemOpen
  \bibfield  {author} {\bibinfo {author} {\bibfnamefont {E.~J.}\ \bibnamefont
  {Mele}},\ }\bibfield  {title} {\enquote {\bibinfo {title} {Band symmetries
  and singularities in twisted multilayer graphene},}\ }\href {\doibase
  10.1103/PhysRevB.84.235439} {\bibfield  {journal} {\bibinfo  {journal} {Phys.
  Rev. B}\ }\textbf {\bibinfo {volume} {84}},\ \bibinfo {pages} {235439}
  (\bibinfo {year} {2011})}\BibitemShut {NoStop}%
\bibitem [{\citenamefont {Park}\ \emph {et~al.}(2015)\citenamefont {Park},
  \citenamefont {Mitchel}, \citenamefont {Elhamri}, \citenamefont {Grazulis},
  \citenamefont {Hoelscher}, \citenamefont {Mahalingam}, \citenamefont {Hwang},
  \citenamefont {Mo},\ and\ \citenamefont {Lee}}]{park2015observation}%
  \BibitemOpen
  \bibfield  {author} {\bibinfo {author} {\bibfnamefont {Jeongho}\ \bibnamefont
  {Park}}, \bibinfo {author} {\bibfnamefont {William~C}\ \bibnamefont
  {Mitchel}}, \bibinfo {author} {\bibfnamefont {Said}\ \bibnamefont {Elhamri}},
  \bibinfo {author} {\bibfnamefont {Lawrence}\ \bibnamefont {Grazulis}},
  \bibinfo {author} {\bibfnamefont {John}\ \bibnamefont {Hoelscher}}, \bibinfo
  {author} {\bibfnamefont {Krishnamurthy}\ \bibnamefont {Mahalingam}}, \bibinfo
  {author} {\bibfnamefont {Choongyu}\ \bibnamefont {Hwang}}, \bibinfo {author}
  {\bibfnamefont {Sung-Kwan}\ \bibnamefont {Mo}}, \ and\ \bibinfo {author}
  {\bibfnamefont {Jonghoon}\ \bibnamefont {Lee}},\ }\bibfield  {title}
  {\enquote {\bibinfo {title} {Observation of the intrinsic bandgap behaviour
  in as-grown epitaxial twisted graphene},}\ }\href@noop {} {\bibfield
  {journal} {\bibinfo  {journal} {Nature communications}\ }\textbf {\bibinfo
  {volume} {6}},\ \bibinfo {pages} {1--8} (\bibinfo {year} {2015})}\BibitemShut
  {NoStop}%
\bibitem [{\citenamefont {Aryasetiawan}\ and\ \citenamefont
  {Gunnarsson}(1998)}]{aryasetiawan1998gw}%
  \BibitemOpen
  \bibfield  {author} {\bibinfo {author} {\bibfnamefont {Ferdi}\ \bibnamefont
  {Aryasetiawan}}\ and\ \bibinfo {author} {\bibfnamefont {Olle}\ \bibnamefont
  {Gunnarsson}},\ }\bibfield  {title} {\enquote {\bibinfo {title} {The GW
  method},}\ }\href@noop {} {\bibfield  {journal} {\bibinfo  {journal} {Reports
  on Progress in Physics}\ }\textbf {\bibinfo {volume} {61}},\ \bibinfo {pages}
  {237} (\bibinfo {year} {1998})}\BibitemShut {NoStop}%
\end{thebibliography}%

\end{document}